\documentclass[conference,compsoc]{IEEEtran}
\def\BibTeX{{\rm B\kern-.05em{\sc i\kern-.025em b}\kern-.08emT\kern-.1667em\lower.7ex\hbox{E}\kern-.125emX}}

\usepackage[dvipsnames]{xcolor}

\usepackage{nicefrac}
\usepackage{array,framed}
\usepackage{booktabs}

\usepackage{textcomp,amssymb}
\usepackage{setspace}

\usepackage{latexsym,fancyhdr,xurl}
\usepackage{enumerate}
\usepackage{algorithm2e}
\usepackage{algpseudocode}
\usepackage{graphics}
\usepackage{xparse} 
\usepackage{xspace}
\usepackage{multirow}
\usepackage{csvsimple}
\usepackage{balance}
\usepackage{csquotes} 
\usepackage{subfigure}
\usepackage{float}
\floatstyle{plaintop}
\restylefloat{table}
\usepackage[tableposition=top]{caption}

\usepackage{
  tikz,
  pgfplots,
  pgfplotstable
}
\usepackage{graphicx,subfigure,mathenv,hyperref}
\usepackage[inline]{enumitem}

\usepackage[framemethod=TikZ]{mdframed}
\newcounter{theo}[section]

\newenvironment{theo}[2][]{
    \refstepcounter{theo}

\begin{mdframed}[]\relax}{
\end{mdframed}}

\ifstrempty{#1}
{\mdfsetup{
    frametitle={
        \tikz[baseline=(current bounding box.east),outer sep=0pt]
        \node[anchor=east,rectangle,fill=blue!20]
        {\strut Instagram Message};}
    }
}{\mdfsetup{
    frametitle={
        \tikz[baseline=(current bounding box.east),outer sep=0pt]
        \node[anchor=east,rectangle,fill=blue!20]
        {\strut Instagram Message};}
    }
}
\mdfsetup{
    innertopmargin=10pt,linecolor=blue!20,
    linewidth=2pt,topline=true,
    frametitleaboveskip=\dimexpr-\ht\strutbox\relax
}

\newcommand{\eg}{e.g.,\xspace}

\newcommand{\etc}{etc.\xspace}
\newcommand{\BfPara}[1]{\vspace{1mm}{\noindent\bf#1.}\xspace\xspace}
\newcommand{\tsref}[1]{\textsection\ref{#1}\xspace}
\definecolor{lightblue}{HTML}{EBFDFE}
\definecolor{lightpink}{HTML}{FFF7F6}
\definecolor{lightgray}{HTML}{D3D3D3}
\usepackage{balance}
\usepackage{hyperref}

\hypersetup{
	pdfpagemode=pagewidth,
	plainpages=false,
	colorlinks,
	urlcolor=blue!70!black,
	linkcolor=red!70!black,
	citecolor=green!70!black,
	bookmarksnumbered
}

\usetikzlibrary{
  shapes.geometric,
  arrows,
  external,
  pgfplots.groupplots,
  matrix
}

\graphicspath{ {images/} }
\usepackage{pifont}  

\newcommand{\htw}{\textsl{HoneyTweet}\xspace}
\usepackage{circledsteps}
\pgfkeys{/csteps/inner color=azure(colorwheel)}

\pgfplotsset{compat=1.9}

\usepackage{mathtools,}

\DeclareMathAlphabet{\mathcal}{OMS}{cmsy}{m}{n}
\DeclareGraphicsExtensions{
    .png,.PNG,
    .pdf,.PDF,
    .jpg,.mps,.jpeg,.jbig2,.jb2,.JPG,.JPEG,.JBIG2,.JB2}

\usepackage{xparse}
\newcommand{\bnm}{\begin{newmath}}
\newcommand{\enm}{\end{newmath}}

\newcommand{\bea}{\begin{eqnarray*}}
\newcommand{\eea}{\end{eqnarray*}}

\newcommand{\bne}{\begin{newequation}}
\newcommand{\ene}{\end{newequation}}

\newcommand{\bal}{\begin{newalign}}
\newcommand{\eal}{\end{newalign}}

\newenvironment{newalign}{\begin{align}
\setlength{\abovedisplayskip}{4pt}
\setlength{\belowdisplayskip}{4pt}
\setlength{\abovedisplayshortskip}{6pt}
\setlength{\belowdisplayshortskip}{6pt} }{\end{align}}

\newenvironment{newmath}{\begin{displaymath}
\setlength{\abovedisplayskip}{4pt}
\setlength{\belowdisplayskip}{4pt}
\setlength{\abovedisplayshortskip}{6pt}
\setlength{\belowdisplayshortskip}{6pt} }{\end{displaymath}}

\newenvironment{newequation}{\begin{equation}
\setlength{\abovedisplayskip}{4pt}
\setlength{\belowdisplayskip}{4pt}
\setlength{\abovedisplayshortskip}{6pt}
\setlength{\belowdisplayshortskip}{6pt} }{\end{equation}}

\newcounter{ctr}

\newcounter{mytable}
\def\mytable{\begin{centering}\refstepcounter{mytable}}
\def\endmytable{\end{centering}}

\newcounter{myfig}
\def\myfig{\begin{centering}\refstepcounter{myfig}}
\def\endmyfig{\end{centering}}

\newlength{\saveparindent}
\newlength{\saveparskip}
\newcommand{\E}{{\rm I\kern-.3em E}}

\renewcommand{\eqref}[1]{\mbox{Equation~(\ref{#1})}}

\def \part {part}

\renewcommand{\paragraph}[1]{\vspace*{6pt}\noindent\textbf{#1}\;}

\def \blackslug{\hbox{\hskip 1pt \vrule width 4pt height 8pt
    depth 1.5pt \hskip 1pt}}
\def \qed{\quad\blackslug\lower 8.5pt\null\par}

\newcounter{mynote}[section]

\newcommand\ignore[1]{}

\newcounter{rcnote}[section]

\newcounter{mrnote}[section]

\newcounter{fknote}[section]

\newcounter{anote}[section]

\DeclareMathSymbol{\mlq}{\mathord}{operators}{``}
\DeclareMathSymbol{\mrq}{\mathord}{operators}{`'}

\newcommand{\rhf}[2]{R_{f, \gamma}}

\DeclareDocumentCommand{\edist}{o o}{
  \ensuremath{
    \IfNoValueTF{#1}{{d}}{{\sf d}(#1,#2)}
  }
}

\newcommand{\olrk}[1]{\ifx\nursymbol#1\else\!\!\mskip4.5mu plus 0.5mu\left(\mskip0.5mu plus0.5mu #1\mskip1.5mu plus0.5mu \right)\fi}

\NewDocumentCommand{\indseq}{ O{1} O{r} }{{#1}\ldots {#2}}

\begin{document}
\fancyhead{}
\newenvironment{bluetext}{\par\color{blue}}{\par}
\newenvironment{redtext}{\par\color{red}}{\par}
\newenvironment{greentext}{\par\color{ForestGreen}}{\par}

\def\thetitle{Conning the Crypto Conman: End-to-End Analysis of Cryptocurrency-based Technical Support Scams}
\title{\thetitle}

\author{
\IEEEauthorblockN{Bhupendra Acharya\textsuperscript{\textsection}}
\IEEEauthorblockA{
\textit{CISPA}\\
bhupendra.acharya@cispa.de}
\and
\IEEEauthorblockN{Muhammad Saad}
\IEEEauthorblockA{
\textit{PayPal Inc.} \\ 
muhsaad@paypal.com
}
\and
\IEEEauthorblockN{Antonio Emanuele Cinà}
\IEEEauthorblockA{
\textit{Università di Genova}\\
antonio.cina@unige.it}
\and 
\IEEEauthorblockN{Lea Schönherr}
\IEEEauthorblockA{
\textit{CISPA} \\
schoenherr@cispa.de} 
\and
\IEEEauthorblockN{Hoang Dai Nguyen} 
\IEEEauthorblockA{
\textit{Louisiana State University} \\
hngu281@lsu.edu}
\and
\IEEEauthorblockN{Adam Oest}
\IEEEauthorblockA{ 
\textit{PayPal Inc.} \\
aoest@paypal.com}
\and  
\IEEEauthorblockN{Phani Vadrevu}
\IEEEauthorblockA{
\textit{Louisiana State University} \\
kvadrevu@lsu.edu}
\and
\IEEEauthorblockN{Thorsten Holz}
\IEEEauthorblockA{
\textit{CISPA} \\
holz@cispa.de}
}
\date{}

\maketitle
\begingroup\renewcommand\thefootnote{\textsection}
\footnotetext{Part of this work contributed to the author's dissertation at the University of New Orleans.}
\endgroup

\begin{abstract}

The mainstream adoption of cryptocurrencies has led to a surge in wallet-related issues reported by ordinary users on social media platforms. In parallel, there is an increase in an emerging fraud trend called \emph{cryptocurrency-based technical support scam}, in which fraudsters offer fake wallet recovery services and target users experiencing wallet-related issues.  

In this paper, we perform a comprehensive study of crypto\-cur\-ren\-cy-based technical support scams. We present an analysis apparatus called \htw to analyze this kind of scam. Through \htw, we lure over 9K scammers by posting 25K fake wallet support tweets (so-called \emph{honey tweets}). We then deploy automated systems to interact with scammers to analyze their {\em modus operandi}. In our experiments, we observe that scammers use Twitter as a starting point for the scam, after which they pivot to other communication channels (\eg email, Instagram, or Telegram) to complete the fraud activity. We track scammers across those communication channels and bait them into revealing their payment methods. Based on the modes of payment, we uncover two categories of scammers that either request secret key phrase submissions from their victims or direct payments to their digital wallets.
Furthermore, we obtain scam confirmation by deploying {\em honey wallet addresses} and validating private key theft. We also collaborate with the prominent payment service provider by sharing scammer data collections. The payment service provider feedback was consistent with our findings, thereby supporting our methodology and results. By consolidating our analysis across various vantage points, we provide an end-to-end scam lifecycle analysis and propose recommendations for scam mitigation.

\end{abstract}

\section{Introduction}
\label{sec:intro}

Over the last few years, cryptocurrencies have witnessed mainstream adoption, leading to an increase in their user base and price. Prominent cryptocurrencies such as Bitcoin and Ethereum represent {\em knowledge-to-money} transfer~\cite{KumaresanMB15, GoldfederBGN17, DemiragC21}, whereby a user with the {\em knowledge} of a private key can spend funds linked to the corresponding public key. In cryptocurrencies, private keys are usually managed by a \emph{digital wallet} whose access is protected by a secret ``key phrase''. A wallet compromise (i.e. losing access to the key phrase associated with the wallet) results in the risk of losing funds linked with that wallet.
 
With the growing adoption of cryptocurrencies as a popular payment instrument, there is an increase in incidents related to wallet theft or compromise ~\cite{metamaskstolenews, trustwalletstolenews, ledgerwalletstolenews}. Moreover, due to the fact that cryptocurrencies are decentralized, there is no central authority that can assist with non-custodial wallet recovery. To make matters worse, scammers can exploit such opportunities to offer fake service support for wallet recovery. As a result, na\"ive users---who are already distressed by losing access to their cryptocurrency wallets---end up losing more money to these fraudsters. 

In this paper, we conduct a comprehensive analysis of the criminal ecosystem involved in such scams from different vantage points.
Our approach is based on three key observations regarding cryptocurrencies and social media. Our first observation is the increasing use of social media for technical support, including support for cryptocurrency-related activities. According to a recent report~\cite{smartInsightsmediacare}, one in three social media users prefers to contact the brand or business support through social media. To accommodate such requests, businesses and brands are ensuring their social media presence by managing customer accounts across various social media platforms (\eg Twitter/X, Instagram, and others). As a result, there is a gradual shift in the mode of technical support from traditional channels (\eg phone calls) to social media platforms. 

Our second observation concerns the abuse of social media by technical support scammers. In a technical support scam, fraudsters trick victims by offering fake customer support for a technical issue. During interactions with the victims, fraudsters try to steal their money by obtaining credentials or charging a (typically high) fee for a service that is subsequently not provided. Technical support scams are commonly conducted through phone calls or emails~\cite{perisci2018, TuD0A19, DerakhshanHB21}. However, realizing the increasing usage of social media for technical support, scammers are also exploiting the opportunity to target victims on social media platforms. 

Our third observation is the use of social media platforms to popularize cryptocurrency-related applications (so-called \emph{Crypto Twitter}), which inevitably contributes to their mass adoption. A recent example is the rapid growth of non-fungible token (NFT) marketplaces that witnessed a sale of \$21 billion in 2021~\cite{howcroft_2022}. As noted in recent work~\cite{KapoorGMYGK22}, social media platforms (especially Twitter) played an instrumental role in boosting NFT trading. Therefore, it can be inferred that social media platforms are serving as a catalyst to amplify the usage and adoption of cryptocurrencies. 

To put our three observations in context: The cryptocurrency ecosystem is evolving fast, and social media platforms enable ordinary users to learn about, adopt, and popularize new cryptocurrency offerings. At the same time, social media platforms are also used by businesses to provide technical support to their customers. Fraudsters are starting to exploit these developments to target social media users with fake technical support scams related to cryptocurrency products. We note that these scams are becoming widespread, as indicated by recent reports from the Federal Trade Commission (FTC)~\cite{ftcsocialalert1, ftcsocialalert2}. Despite growing concerns about cryptocurrency-related technical support scams, there is limited prior work that analyzes these emerging scam trends on social media~\cite{xigao2023doublenothing, 9493255, phillips2020tracing}. In particular, there is no research that conducts a comprehensive analysis of cryptocurrency-based technical support scams.

In this paper, we close this gap by proposing a new scam detection apparatus called \htw that uncovers crypto\-cur\-ren\-cy-based technical support scams on Twitter. In a nutshell, \htw applies a {\em Conning the Conman} approach by posting unique and automated tweets (so-called \emph{honey tweets}) with technical support request keywords to set traps for scammers. After luring the scammers, we deploy engagement tools that follow the instructions provided by scammers and bait them into revealing their payment information. As a result of this systematic interaction through \htw, we (1) identify different types of technical support scammers, (2) uncover their {\em modus operandi}, and (3) profile them based on their profile footprints across different social media and payment platforms. Through \htw, we lure over 9K scammers in three months and track their footprints across six social media platforms. We validate our findings by sharing information with PayPal, and after receiving scam confirmation, we further disclose our findings to the social media platforms where scammers target cryptocurrency users.

\BfPara{Contributions} Our key proposition is \htw: a scam detection method that employs clever techniques to uncover emerging fraud trends and engage with scammers to perform a scam lifecycle analysis. We deploy \htw to study cryptocurrency-based technical support scams, and our findings are summarized below as key contributions.

\begin{enumerate}[leftmargin=*]
    \item \textbf{Scam Detection} Through \htw, we generate tweets tailored to lure scammers on Twitter. We deploy \htw from October 2022 to January 2023 and post 25K automated tweets from test accounts. Our tweets lured over 9K technical support scammers. 

    \item \textbf{Scammer Profiling.} After collecting 9K scammer profiles, we collect Twitter profile features and apply standard machine learning models to study affinities among scammer profiles. Our clustering analysis indicates that scammers often try to masquerade popular cryptocurrency exchanges as well as NFT tokens to lure victims.

    \item \textbf{Scam Lifecycle Analysis.} We apply automated and manual techniques to interact with scammers and follow their instructions to study the end-to-end scam lifecycle. We observe that scammers use Twitter as a starting point for the fraud and subsequently pivot to other social media platforms where the rest of the fraud activity is conducted. In our interactions with scammers, we bait them into revealing their payment information. We uncover two broad categories of scammers based on their payment choices. Scammers in the first category request private key submissions, while scammers in the second category request payments to their cryptocurrency addresses or digital accounts.  

    \item \textbf{Scam Validation.} We perform scam validation by setting up {\em honey wallet addresses} and conducting experiments to observe private key theft by scammers. Our analysis of the observed Bitcoin and Ethereum addresses shows that the scams are successful in practice. We also collaborated with PayPal and shared information about scammers requesting PayPal payments. PayPal confirmed our findings about fraud conducted by those accounts, thereby validating the efficacy of \htw in scam detection.
 
\end{enumerate}

For fostering future research, we publish the code under an open-source license~\cite{honeytweetcode}. 
For data protection reasons, we will only share data on scammers to interested academics or entities on request.

\BfPara{Responsible Disclosure and Ethical Considerations} Since we observed cross-platform interactions with the same set of scammers, we treat this scam to be an ecosystem problem. Therefore, we have disclosed our findings to prominent social media platforms including Google, Twitter, Instagram, Telegram, and WhatsApp. Our experiments involved a deception-based user study in which we omitted the debriefing and after-study consent procedures to protect the authors from retribution attacks. The IRBs at institutions supporting this work agreed with this methodology.
    
\section{HoneyTweet: System Anatomy}
\label{sec:system-description}

\begin{figure}[tb]
\centering
\includegraphics[width=.38\textwidth]{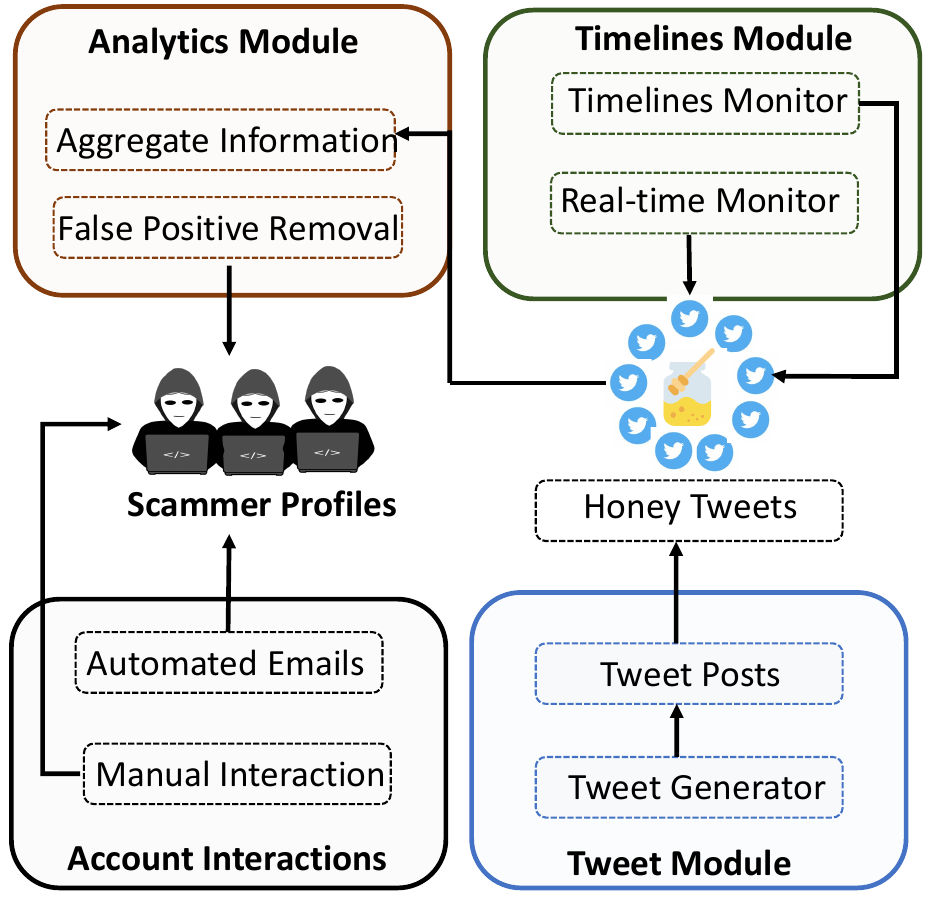}\hfill
\caption{Overview of the \htw data collection and analysis pipeline. Our workflow consists of three modules (tweet, analytics, and timelines module). Additionally, we implemented an account interaction component where we performed automated and manual interactions with scammers outside of the Twitter platform (more details in Sec.~\ref{sec:pivoting}). This figure displays the anatomy of tweet posts that we perform to lure scammers.}
\label{fig:crypto_sys_design_active}
\end{figure}

In this section, we describe the data collection and analysis process underlying \htw. 
From a high-level point of view, \htw consists of three modules: (1) a module that generates crypto-related honey tweets, (2) a timelines module that collects additional information about scammers beyond their interaction with our honey tweets, and (3) an analytics module that enriches the collected information to enable a quantitative data assessment (\eg, to study fraud mechanics, the scam lifecycle, and affinities among fraudster profiles). \autoref{fig:crypto_sys_design_active} provides a general overview of our data collection and analysis workflow, and we describe the implementation aspects of the three key modules below. 

\subsection{Tweets Module}
\label{ssec:sys_active_tweets}
\begin{figure}[tb]
\centering
\includegraphics[width=.45\textwidth]{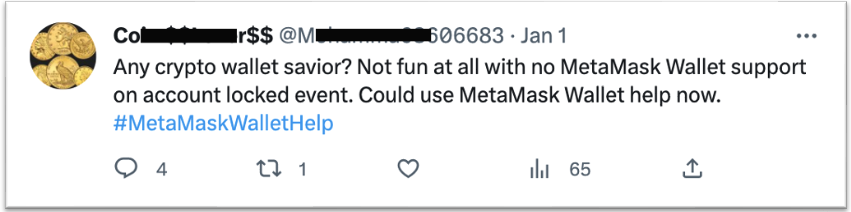}\hfill
\caption{A sample tweet from our \textsf{Tweet Generator} module. Note that the profile setup reflects a cryptocurrency enthusiast. The first sentence is a greeting, followed by two sentences indicating the problem and requesting help. We also included hashtags to enhance the tweet visibility.}
\label{fig:sample_tweet}
\end{figure}

The first module is the \textsf{tweets module}, which generates honey tweets for scammers and stores them in the in-house MongoDB. The module uses a \textsf{honey tweet generator} component to generate tweet posts consisting of three sentences. The first sentence contains a greeting such as~\emph{``Hey there!''} or~\emph{``Hi, Wallet Support!''}. The second sentence consists of a sequence of words that serve as bait for the scammers. For instance, the second sentence contains \emph{``wallet name''} and keywords such as \emph{``help''} or \emph{``support''} with frustrated or account-blocked sentiments. The third sentence adds a sense of urgency to the request by using keywords or hashtags such as \emph{``any references asap please?''}, \emph{``I am in dire need!''}, and \emph{``\#walletnamesupport''}. In~\autoref{fig:sample_tweet}, we provide a sample tweet generated by the \textsf{tweets module}. We also applied a random generator function in the \textsf{tweets module} to select a random wallet from 10 popular wallet providers. We explain the rationale for our \htw design in Appendix~\ref{sec:honeytweet_design_rationale}. More details on wallet providers can be found in~\autoref{table:wallet_target_breakdown}. 

To ensure that our tweets mimic legitimate requests, we apply two techniques. First, we apply filters to prevent duplicate tweet posts. After crafting a random tweet, our \textsf{honey tweet generator} cross-references that tweet to our database to ensure that no previous tweet with the same content has been posted before. Next, we check the 280-character limit to check that the tweet draft conforms to the platform's specifications. 

After generating a series of tweets, the \textsf{honey tweet generator} posts them on our honey accounts using Twitter's \emph{Update API}. The posts were generated at 15-minute intervals. Our rationale for selecting the 15-minute window was based on the initial system incubating period. We performed a manual experiment for two days and posted 30 honey tweets on our test accounts. We observed that scammers roughly took 10-15 minutes to respond to our tweets. As a result, we set a 15-minute interval between tweet postings. 

We set up four Twitter accounts through which we posted honey tweets. Our rationale for selecting four accounts is as follows. We needed a setup with a fail-safe mechanism in case one or two profiles were suspended by the platform or detected by scammers. To prevent platform suspension, we proactively provided all experiment details to the Twitter Developer Platform. Moreover, having more than one account enabled us to collect a sizable number of scammer profiles that replied to our tweets. Finally, we did not want to spam the platform by setting up a large number of accounts that continuously generated honey tweets. Therefore, to ensure an optimal balance between the two conditions, we set up four accounts. We configured each account's profile features such as name, description, and profile image, to mimic cryptocurrency enthusiasts. Moreover, through account APIs, we posted automated tweets as well as collected scammer profiles.

\subsection{Timelines Module}
\label{ssec:timelines_module}

The second module is the~\textsf{timelines module}, which receives a list of scammer profiles from the~\textsf{tweets module} and then crawls their timeline using Twitter APIs. The key idea behind the \textsf{timelines module} is to collect additional information on scammers beyond their interaction with our honey tweets. Moreover, the module maintains a timestamped record of the scammers' profiles in case they change information or their account is suspended. In the following, we describe the breakdown of the timeline monitor, which performs crawls of scammers' profiles on a daily basis.  

\BfPara{Scammer Profile Details}
For the profile details, we use the \emph{user detail} Twitter API~\cite{twitteruserdetailapi}. In particular, we collect features including name, location, description, followers count, following count, and profile picture. We also download the profile pictures to track profile picture changes during the account's lifetime. 

\BfPara{Scammer Posts}
For tracking the scammers' daily interactions with the potential victim's account, we also used the \emph{user timeline API}, which returns the associated tweets from the specified user account \cite{twitterusertimelineapi}. The tweets include replies, quoted tweets, and retweets. For each data fetch, we apply a marking point pull to avoid duplicate pulls for a given scammer profile.

\subsection{Analysis Module}
\label{ssec:analysis_module}

The third module is the~\textsf{analysis module}, which analyzes the data collected by the~\textsf{timelines module} and extracts useful attributes. The~\textsf{analysis module} performs a quantitative data assessment to analyze the scam mechanics, the scam lifecycle, and the affinities between scam profiles. 

To preserve data quality, the \textsf{analysis module} also applies filters to remove false positives. For example, it could be possible that official cryptocurrency exchanges respond to our honey tweets, and including such exchanges in the scam dataset could contaminate our analysis. To avoid such false positives, we developed a Selenium-based web scrapper tool that collects Twitter accounts of official cryptocurrency exchanges. In total, we collected data on 256 exchanges and 8,538 cryptocurrencies. We obtained the list of exchanges and cryptocurrencies from CoinMarketCap~\cite{coinmarketcapall}. 

We also exclude verified accounts from our analysis. Generally, verified accounts are less likely to engage in such scams because Twitter uses real-world attributes of verified accounts. We acknowledge that this assessment may no longer be valid, as Twitter recently began offering paid subscriptions. However, we leave this analysis for future work and conservatively removed verified accounts from our data set. Additionally, we also filtered benign users that interacted with our tweets. The key distinction between benign users and scammers is that benign users do not provide any information that enables further interaction with them. In contrast, scammers respond with an archetypal message with clear details about contacting them (often outside of Twitter, see Sec.~\tsref{sec:pivoting}). Therefore, a clear distinction between benign and scam accounts is that benign accounts do not attempt further engagements in their replies. On the other hand, scammers provide clear details to contact them so that the victim can be lured into a trap. We acknowledge that our process for removing potential false positives is rather conservative, as we might have excluded several scam accounts in the process. However, this conservative filtering ensures that the population we examined contained only scam accounts.  

\begin{figure}[tb]
\centering
\includegraphics[width=.45\textwidth]{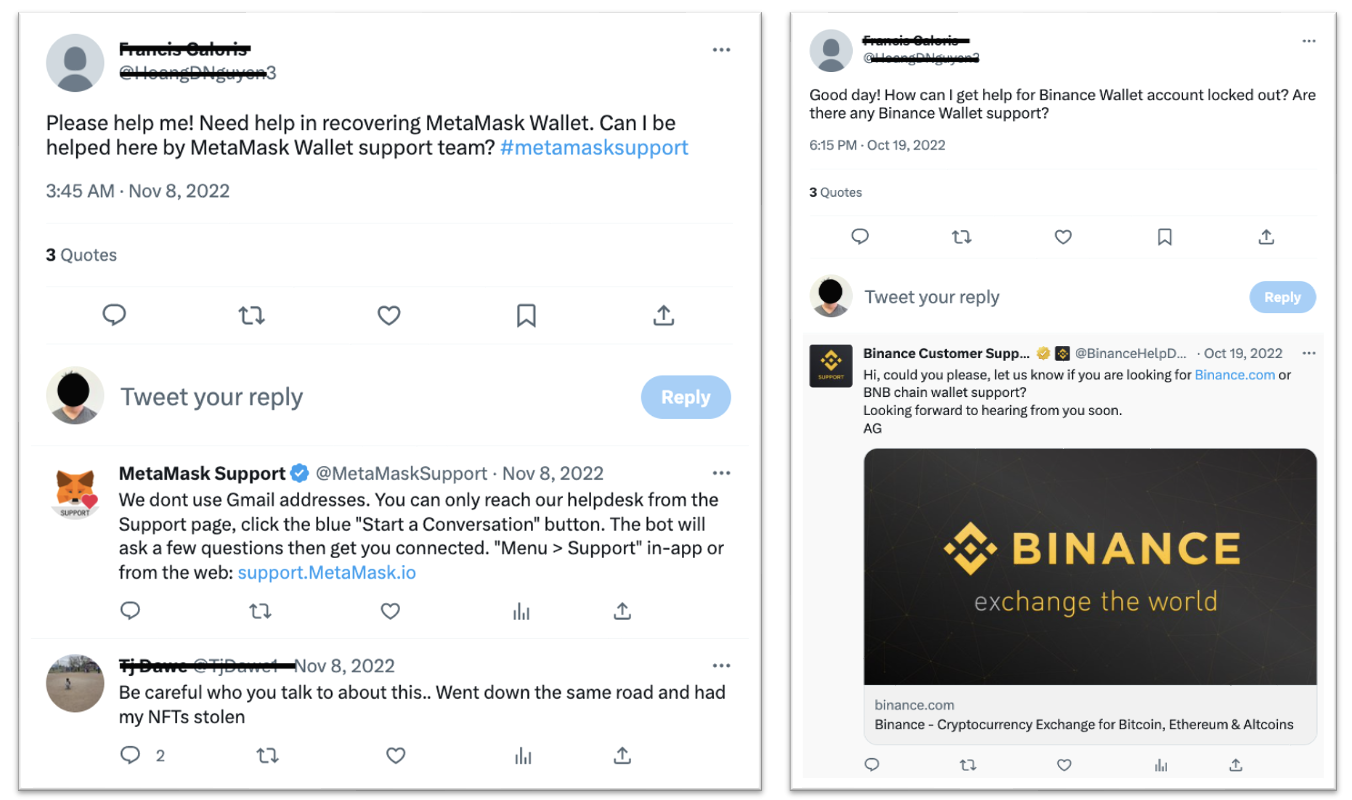}\hfill
 \caption{False Positive examples of benign interaction performed by official wallet support and regular Internet users. The left image shows two replies, one from the official MetaMask support and the second from regular Internet interaction. The right graph displays Binance's official support link to the page and offers support for wallet issues.} 
\label{fig:fp_screen_shot}
\end{figure}

Via the filtering process outlined above, we identified 44 accounts that might potentially be false positives. Among them, four accounts belonged to official support channels (see \autoref{fig:fp_screen_shot} for an example), two were verified users, and 40 were potentially benign accounts that did not offer wallet recovery support or any information to contact them. We excluded these accounts from further analysis. We provide additional details on tweet filtering in Appendix~\ref{sec:tweet_filtering}.
     
\section{Technical Scams: Twitter Interactions}
\label{sec:dataset_analysis}
In this section, we provide details about scammer interactions with our honey tweets. More specifically, we (1) showcase the interaction types used by scammers to interact with their victims and (2) conduct a deep dive to dissect the main characteristics of scammer profiles. Our analysis in this section is focused on all engagements {\em observable} on Twitter. Later in Sec.~\tsref{sec:pivoting}, we will further elaborate on interactions with scammers \emph{outside} of Twitter. 

\begin{table}[b]
\centering
\scalebox{0.69}{
\renewcommand{\arraystretch}{1.3}
\setlength\tabcolsep{3pt} 
\begin{tabular}{lcccccc}
\toprule
\bf{Modes of} & \bf{Total } & \bf{Distinct} & \bf{Distinct} & \bf{Suspended} & \bf{Inactive} & \bf{All} \\
\bf{Interaction} & \bf{\#} & \bf{\#} & \bf{TweetID \#} & \bf{Account \#} & \bf{Account \#} & \bf{Account \#} \\
\midrule
Replies & 28424	& 19104	& 9460	& 5109	& 7521 & 7953 \\
Retweets & 1491 & 1491 & 1333 & 293 & 394 & 452\\
Quoted Tweets & 10218 & 10212 & 5202 & 1517	& 2124 & 2434\\
Likes & 15901 & 15901 & 7568 & 2673	& 3935 & 4265 \\
Follow & 660 & 660 & 9968 & 398 & 588 & 660\\
Total Interact & 57226 & 47368 & 20740 & 6423 & 8954 & 9149\\
\bottomrule
\caption{Break down Modes of tweet interaction by scammers with \emph{HoneyTweet}.}
\label{table:over_all_tweet_interact}
\end{tabular} }

\end{table}

\subsection{Modes of Scammer Interactions}
\label{ssec:scammer_ineraction_modes}

\begin{figure*}[tb]
  \centering
\begin{subfigure}[b] [Scammer Interactions\label{fig:si_at}]{\includegraphics[width=0.32\textwidth]{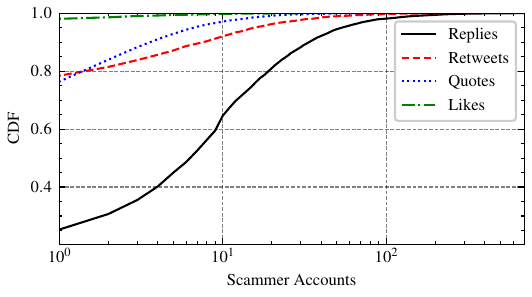}} 
\end{subfigure}
   \hfill
\begin{subfigure}[b][Following Count\label{fig:si_af}]{\includegraphics[width=0.32\textwidth]{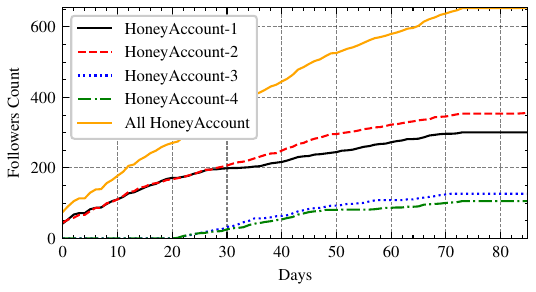}}
\end{subfigure}
   \hfill
\begin{subfigure}[b][Cumulative Number of Scammers\label{fig:si_sd}]{\includegraphics[width=0.32\textwidth]{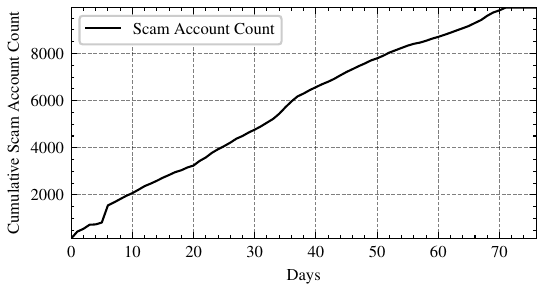}}
\end{subfigure}
\caption{Interactions of scammers with \htw. Figure~\ref{fig:si_at} shows the overall interaction in the form of replies, retweets, quoted tweets, and likes seen over the experiment duration. \autoref{fig:si_af} shows the {\em following} count of each honey tweet account. \autoref{fig:si_sd} shows the cumulative sum of scammers based on their total interactions with \htw. }
\label{fig:scammer_interaction_with_honey_tweet}
\end{figure*}

The dataset collected for our analysis consists of 25K honey tweets and 9,149 scammer profiles collected between October 14, 2022, and January 02, 2023. In~\autoref{table:over_all_tweet_interact}, we provide a high-level summary of engagement mechanics used by scammers to interact with \htw. The first column in~\autoref{table:over_all_tweet_interact} shows the different ways fraudsters interacted with our tweets. The subsequent three columns show the total count, the distinct count, and the unique \emph{TweetID} corresponding to each mode of interaction. Moreover, we observed that Twitter also suspended scam accounts for violations of platform policies. The number of suspended accounts linked to each mode of interaction is shown in the fifth column. Finally, the sixth column reports the number of accounts that became inactive, which includes suspended accounts as well as deactivated accounts (due to account deletion) as of January 16, 2023 (two weeks after the end of our data collection period). Overall, 20,740 (79.42\%)  of the total 25K tweets received \emph{at least one} interaction from scammer accounts. In total, 9,149 scammer accounts interacted with 20,740 tweets, which is an indication of the popularity of this scam scheme. Interestingly, as many as 8,954 of these accounts (97.86\%) ended up becoming {\em marked as inactive}, thus showing the highly ephemeral nature (largely due to suspensions) of the accounts that interacted with our honey tweets. See Appendix~\ref{sec:evaluation} for more discussion on the ground truth of our dataset.

In~\autoref{fig:scammer_interaction_with_honey_tweet}, we show the overall interaction between scammers and \htw to provide a high-level overview of our dataset. Figure~\ref{fig:si_at} presents the CDF of scammer interactions with \htw in terms of \emph{replies}, \emph{quoted tweets}, \emph{retweets}, and \emph{likes}. The total number of scammer accounts that followed our four honey tweet profiles is shown in~\autoref{fig:si_af}. Finally,~\autoref{fig:si_sd} presents the total number of scammer accounts that we identified during the experiment duration. Note that all accounts in~\autoref{fig:si_sd} are calculated based on all modes of interaction in~\autoref{table:over_all_tweet_interact}. With these high-level statistics in mind, we next discuss each mode of interaction to uncover the popular means that scammers use to bait victims on Twitter. 

\BfPara{Replies} Our results show that 7,953 (86.92\%) of the scammers preferred to interact with our tweets through replies. We found that these accounts sent 284,424 replies to 9,460 (45.61\%) distinct tweets that we posted. This observation suggests that tweet reply is the most popular interaction instrument through which scammers engage with potential victims on Twitter. Of the 28,424 replies in 9,460 distinct honey tweets, we found 19,104 distinct texts. Thus, we also observed that scammers performed repeated text in 9,320/28,424 (32.78\%) distinct tweets. We suspect that the repeated text replies likely indicate scripted behavior.
  
\BfPara{Retweets} Our dataset revealed that retweets were the least popular mode of interaction. Only 452 (0.5\%) scammers retweeted 1,333 (6.4\%) of distinct honey tweets. A likely reason for the low retweet count is that retweets do not lead to a conversation between the fraudster and the victim. By retweeting, the scammers merely relay the victim's tweet to their followers. At best, retweeting increases the victim's visibility to other potential scammers.  

\BfPara{Quoted Tweets} Roughly speaking, quoted tweets combine the functionality of replies and retweets. Through quoted tweets, users can generate a response to the original tweet and display that response to their followers. We found that among the total 20,740 tweets that received an interaction, 5,202 (25.08\%) tweets were quoted by 2,434 (25.08\%) scammer accounts. We also found that scammers who quoted tweets accounted for the lowest number of suspended accounts.

\BfPara{Likes} On Twitter, likes (indicated by a heart icon) signify that someone appreciated or showed concern regarding the tweet. Our dataset revealed 2,673 (41.61\%) of suspended scamming accounts \emph{liked} 3,678 (17.73\%) distinct tweets. We also found that after generating a like, scammers typically send a reply or quote the tweet. 

\BfPara{Follows} Among all the scammer accounts, 660 accounts followed our honey tweet profiles. Among those 660 accounts, 398 accounts were subsequently suspended by Twitter. After following, the scammer accounts interacted with 9,968/20,740 (48.06\%) of the total honey tweets posted on each profile. We presume that by following victim accounts, scammers intend to monitor victims and hope to receive a follow-back that can enable a private conversation between them through the {\em Direct Messaging} functionality of Twitter. 

\BfPara{Key Takeaways} In summary, we detected 9K scammer accounts that interacted with \htw. By analyzing their modes of interaction, we derive the following three conclusions: (1) Scammers use various means to interact with their victims. (2) The most preferred way of engagement is replying to tweets, followed by quoted tweets. (3) Twitter has put guard rails in place to suspend scammer accounts. However, 2,726 scammer accounts escaped suspension by bypassing Twitter's detection systems.

It is worth noting that the risk of account suspension leaves scammers with a limited time window to trap potential victims. Therefore, we suspect that scammers use Twitter as a starting point to engage their victims. Subsequently, scammers could likely pivot from Twitter to other communication channels where the risk of account suspensions is low. In Sec.~\tsref{sec:pivoting}, we will provide empirical evidence that scammers invite their victims to other communication channels to complete the scam. Before we present this analysis, we take a closer look at the profile-based attributes of scammers, which we present below.

\subsection{Analysis of Scammer Profiles}
\label{sec:scammer_timelines}

Next, we analyze the key attributes of scammer profiles, including user details, lifespan, and timeline interactions. Based on our analysis, we will derive affinities among scammers. 
In the first step, we use the \textit{timelines module} (see \autoref{fig:crypto_sys_design_active}) to analyze the account lifespan, tweet source, language preference, and geographical distribution to analyze the user details. In the following, we provide details on each attribute. 

\BfPara{\textbf{Account Lifespan}} Our results show that out of 9,149 scam accounts, 5,423 (59.27\%) accounts were created in the year 2022. The remaining 3,726 (40.73\%) accounts were created between 2009 and 2021. We also analyzed \emph{user details} API errors to study the account suspensions. We found 6,423 accounts with \emph{Forbidden} API error, indicating that these accounts were suspended by Twitter due to policy violations. We also found 2,531 accounts that returned a \emph{Not Found} API error, indicating that the accounts were deleted or deactivated. 

\begin{table}[b]
\centering
\setlength\tabcolsep{3.5pt} 
\scriptsize
\begin{tabular}{lcc}
\toprule
\bf{Source} & \bf{TweetID Interact \% } & \bf{Scammers \#}  \\
\midrule
Twitter for iPhone & 71.42 & 6520\\
Twitter for Android &  66.01 & 6620\\
Twitter for Web App & 13.27 & 2822\\
Twitter Deck & 0.42 & 196\\
Twitter for iPad & 0.11 & 44\\
\bottomrule
\end{tabular}
\caption{Overview of Tweet sources. Our results indicate that scammers prefer mobile devices to interact with the victims. }
\label{table:scammer_visitor_platform}
\end{table}

\BfPara{\textbf{Tweet Sources}} An analysis of the tweet sources can highlight the popular platforms and devices used by scammers. In our dataset, we found that 99\% of the fraudster accounts tweeted from three main sources, namely iPhone, Android, and Web App. Fewer than 1\% of the accounts used Twitter through Deck~\cite{twitterdeck} and iPad. In~\autoref{table:scammer_visitor_platform}, we provide an overview of the tweet sources through which scammers interacted with \htw across all modes of interactions.~\autoref{table:scammer_visitor_platform} shows that iPhone is the most common source through which scammers interacted with our honey tweets, followed by Android and Web App, respectively.  Moreover, we found that scammers frequently switched their devices. Out of 9,149 scam accounts, 6243 (68.23\%) accounts used more than one source to interact with \htw. Among them, 5,773 (63.09\%) accounts used both iPhone and Android, while 2,089 accounts used all three popular sources.

\BfPara{\textbf{Tweet Language Preference}} Our data shows that 97.52\% of the scam accounts preferred English (EN) as their language. 1.86\% of accounts used French, while the remaining accounts used different languages, including Indian (IN), Japanese (JP), Chinese (ZH), Spanish (ES), Welsh (CY), and Lithuanian (LT). 

\BfPara{\textbf{Geographical visit}} We found that 65.04\% of scam accounts did not provide the account location. It is plausible to expect such behavior, as fraudsters would naturally avoid disclosing their location. However, among the accounts that revealed their locations, 10.46\% were linked to the USA, and the remaining 24.50\% were distributed across different geographic locations such as the UK, France, Nigeria, Canada, India, and Congo.  
We also found non-geolocations, including Cryptoverse, Earth, Blockchain, Metaverse, etc. 

\BfPara{Key Takeaways} Based on our analysis, the key takeaways are that scam accounts prefer to interact through mobile devices like iPhone and Android. Moreover, since English is the most commonly used language on Twitter, scammers set their preferred language as English in order to maximally trap potential victims. Finally, scam accounts commonly avoid revealing their geolocations, likely to evade tracking. 

\subsection{Scammer Profile Pictures}~\label{sec:scammer_profile_pictures}
In this section, we analyze the profile pictures of scammer accounts and use unsupervised clustering to study affinities between these pictures. For that purpose, we collected all profile pictures and used a pre-trained visual model called CLIP~\cite{Radford21CLIP} for feature extraction. For each profile picture, we extracted the CLIP token embedding, re-scaled them at $224\times224$ pixels resolution, and visualized it using Uniform Manifold Approximation and Projection (UMAP)~\cite{McInnes2018}. We use standard clustering algorithms, namely HDBSCAN~\cite{McInnes17hdbscan} and single-linkage hierarchical clustering~\cite{Hastie01Hierarchical}, to group pictures and remove anomalies. To enhance the quality of our clustering pipeline, we carefully select appropriate hyperparameters through extensive experimentation and validation. Detailed information on these choices can be found in Appendix~\ref{sec:hyperparameters}.

 \begin{figure*}[t]
    \includegraphics[width=1.0\textwidth]{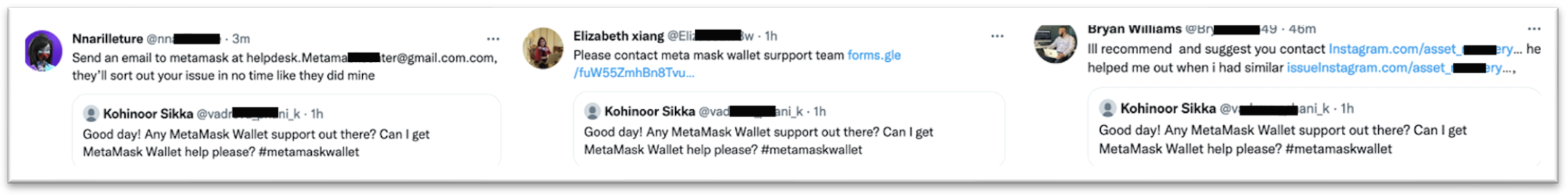}\hfill
            \caption{Examples of scam accounts asking potential victims to connect through email, Google forms, and Instagram. These interactions indicate that scammers use Twitter as a starting point for fraud before pivoting to other platforms. } 
         \label{fig:scam_media}
\end{figure*}

\BfPara{Clustering Results}
With our unsupervised clustering approach, we identified seven groups of profile pictures commonly used by scammers. The seven groups included: \textit{NFTs}, \textit{Male}, \textit{Female}, \textit{Tech Support}, \textit{Wallets}, \textit{Default Twitter Profile Image}, and \textit{Miscellaneous}. In~\autoref{table:scammer_clusters_new}, we showcase the output of our clustering model. We also pair the output with the number of scammer profiles and their modes of interaction. Our results show that scammers commonly use NFTs as their profile pictures. NFT clusters typically included  CryptoPunks, Bored Ape, and Yacht Club. We observed that scammers also use celebrity profile pictures in the \emph{Male} and \emph{Female} clusters. The \emph{Tech Support} cluster contained images of computer desks, credential recovery logos, and hacker silhouettes. Furthermore, we found that scammers from the \emph{Wallet} cluster specifically targeted prominent exchanges by displaying their logos. Those exchanges included Coinbase, Trust Wallet, BitPay, and Badger. We also identified accounts with default Twitter profile images. Finally, the \emph{Miscellaneous} cluster contained pictures of animals, cropped logos of cryptocurrency wallets, or cartoon images. In~\autoref{fig:scammers_cluster_1} and~\autoref{fig:scammers_cluster_2} (see Appendix~\ref{sec:profile_clustering}), we present example images from each main category.  

\begin{table}[tb]
\resizebox{0.49\textwidth}{!}{
  \renewcommand{\arraystretch}{1.2}
  \setlength\tabcolsep{2.5pt} 
  \begin{tabular}{lrrrrr}
\toprule
\begin{tabular}[c]{@{}l@{}}\textbf{Cluster}\\ \textbf{Label}\end{tabular} & \textbf{Scammer \%} & \textbf{Followers \%} & \textbf{Replies \%} & \begin{tabular}[c]{@{}l@{}}\textbf{Quoted} \\ \textbf{Tweets \%} \end{tabular} & \textbf{Suspended \%}  \\
\toprule
NFTs                                                    & 22.98        & 20.67         & 20.95        & 17.18    & 70.91          \\ 
Male                                                    & 22.81        & 22.81         & 22.56         & 30.86     & 64.53          \\
Female                                                  & 22.56        & 19.07         & 25.20         & 28.78      & 66.22             \\ 
Tech Support                                            & 11.87        & 8.91          & 10.19         & 8.30        & 55.54            \\
Wallets                                                 & 11.64        & 23.35         & 9.89          & 4.72     & 55.79              \\
Default Image                                           &7.01           & 2.85          & 9.10          & 8.29     & 58.68             \\
Miscellaneous                                           & 1.09         & 2.31          & 2.07           & 1.83    & 59.0 \\ \bottomrule                                        
\end{tabular}
}
\caption{Results obtained through the clustering analysis. We note that scammers most commonly use popular NFT images as profile pictures.}
\label{table:scammer_clusters_new}
\end{table}

\BfPara{Cluster Engagements} Our results in~\autoref{table:scammer_clusters_new} show that accounts from \emph{NFTs}, \emph{Male}, and \emph{Female} clusters accounted for 68.35\% of the total scam accounts. Combined, the three clusters accounted for 68.71\%  and 76.82\% of the total replies and quoted tweets, respectively. Although there were only half as many accounts in the \emph{Wallet} cluster as in the \emph{NFTs/Male/Female} clusters, their following count for our honey profiles was greater than all other clusters. We hypothesize that by following accounts, scammers may try to reduce the risk of Twitter suspension. Generally, if scammers receive a follow-back from the victim, it enables them to gain credibility and also send direct messages to the victim through Twitter. Our intuition is supported by data in~\autoref{table:scammer_clusters_new}, where we observed that the percentage of accounts in the wallet cluster generally experienced lower platform suspension. 

\BfPara{Key Takeaways} In summary, our clustering analysis provides deeper insights into the scammer profiles, specifically their choice of selecting display pictures. We observed that scammers use NFT pictures or clone exchange logos to appear as legitimate technical service providers. Since NFTs have gained significant popularity in the last year, we can expect an increase in the technical issues related to the MetaMask wallet. It makes sense that scammers also choose NFT images as display pictures to gain the attention of users with NFT-related technical issues.

{\color{blue}}
\subsection{\textbf{Tweet Content Analysis}} 

We now present the content analysis of tweets posted by scammers. We observed that in response to a user's help request, scammers typically posted a reply with a URL. Naturally, a URL embedded with tweet indicates that the tweet author expects the reader to click on the URL and read its contents. In our context, such tweets with URLs indicated how scammers aimed to (mis)guide users toward the next step in fake wallet recovery. Note that when a user posts a tweet containing a URL, Twitter populates the URL metadata underneath the tweet contents, thereby revealing insights into the URL contents. We leveraged this functionality to collect those URLs and categorize them based on their hosted content. In other words, content analysis revealed the website or online platform where scammers further engage with their victims.

Our results showed that URLs posted by scammers frequently navigated users to either Twitter Direct Messages or communication channels outside of Twitter including Instagram, WhatsApp, Telegram \etc In~\autoref{fig:scam_media}, we provide a sample interaction that shows scammer pivoting from Twitter to other communication channels. In situations where we did not find URLs in tweet contents, we extracted the communication channel based on regular expression matching the email address. In~\autoref{table:scammer_scam_media_comparison_table}, we provide the distribution of those communication channels. We categorize their distribution across replies received by our honey profiles and total replies sent to all potential victims with whom the scammers interacted through Twitter. We found that scammers referred 6,167 unique communication channels to a total of 194,363 Twitter users. Among those 6,167 communication channels, 712 were email addresses through which scammers further engaged with their victims. Among 712 unique email addresses shared by scammers, 375 addresses were shared with our honey profiles. 

\BfPara{Key Takeaways}
From tweet content analysis, we derived the following conclusions: Scammers embed URLs in replies to redirect users to other communication channels. This activity shows that Twitter serves as a starting point for the scam, and the fraud activity completes on a different communication channel. Among those channels, we observed that scammers prefer to redirect users to email-based platforms. We suspect that scammers pivot from Twitter to email-based platforms in order to continue the fraud activity even if their Twitter accounts are suspended. Therefore, to fully understand the scam lifecycle, it is pertinent to study scammer interactions on other communication channels outside of Twitter. In the next section, we present our experiments through which we study the three popular communication channels used by scammers. 

Additionally, we also conducted account clustering based on each of the shared e-platform identifiers (e-mail and Instagram addresses, form URLs). Our analysis can be found in Appendix~\ref{sec:clustering_scammer_channel}.

\begin{table}[t]
\centering
\setlength\tabcolsep{10pt} 
\renewcommand{\arraystretch}{1.}
\scriptsize
\begin{tabular}{lrr}
\toprule
\bf{Channels} & \bf{Honey Profiles } & \bf{Total}  \\
\midrule
Email & 375 & 712\\
Forms & 1076 & 1995\\
Instagram & 551 & 874\\
Telegram & 259 & 1041\\
Twitter DM & 731 & 919\\
WhatsApp & 106 & 428\\
\midrule
All & 3098 & 6167\\
\bottomrule
\end{tabular}
\caption{Distribution of communication channels posted by scam accounts in our honey tweets posts. Each channel is categorized based on replies sent to honey profiles and all potential victims.}
\label{table:scammer_scam_media_comparison_table}
\end{table}

\section{Technical Scams: Platform Pivoting}
\label{sec:pivoting}

To fully understand the scam lifecycle, we continued our interactions with scammers outside of Twitter by conducting two experiments. Our experiments involved automated and manual interactions with scammers through email and Instagram. Below, we discuss each experiment in detail. 

\subsection{Interactions via Email} \label{sec:email_intreactions}

In \htw, we generated tweets requesting support for various cryptocurrency wallet types. The intuition behind this approach was to gain a holistic understanding of the threat landscape as well as capture the maximum number of scammers that precisely target specific wallet users.~\autoref{table:wallet_target_breakdown} provides a breakdown of wallet addresses along with communication channels requested by scammers to contact them.

\begin{table}[t]
\centering
\scalebox{0.7}{
\setlength\tabcolsep{2.5pt} 
\renewcommand{\arraystretch}{1.2}
\begin{tabular}{lrrrrrrrr}
\toprule
\bf{Wallet} & \bf{Dist. } & \bf{Dist.} & \bf{Dist.} & \bf{Dist.} & \bf{Dist.} & \bf{Dist.} & \bf{Dist.} \\
\bf{Names} & \bf{Email} & \bf{Form} & \bf{Instagram} & \bf{Telegram} & \bf{Twitter DM} & \bf{WhatsApp}  & \bf{All} \\
\midrule
Badger & 51 & 53 & 93 & 25 & 19  & 21 & 262 \\
Binance  & 47 & 64 & 169 & 118 & 40 & 24 & 462 \\
BitPay  & 45 & 71 & 95 & 32 & 24 & 27 & 294 \\
Coinbase  & 62 & 323 & 148 & 83 & 212 & 24 & 852 \\
Exodus  & 50 & 94 & 135 & 28 & 28 & 22 & 357 \\
Free  & 39 & 67 & 84 & 19 & 22 & 17 & 248 \\
Ledger  & 35 & 115 & 93 & 27 & 127 & 16 & 413 \\
MetaMask  & 261 & 312 & 114 & 52 & 73 & 14 & 826 \\
Trezor  & 47 & 61 & 110 & 33 & 28 & 20 & 299 \\
Trust Wallet  & 69 & 497 & 123 & 56 & 162 & 19 & 926 \\
\midrule
Total & 375 & 1076 & 551 & 259 & 731 & 106 & 3098 \\
\bottomrule
\end{tabular}}
\caption{Distribution of responses received for different cryptocurrency wallet types with regard to different communication channels through which scammers expected us to communicate outside of Twitter. }
\label{table:wallet_target_breakdown}
\end{table}

\begin{figure}[t]
\centering
\includegraphics[width=.48\textwidth]{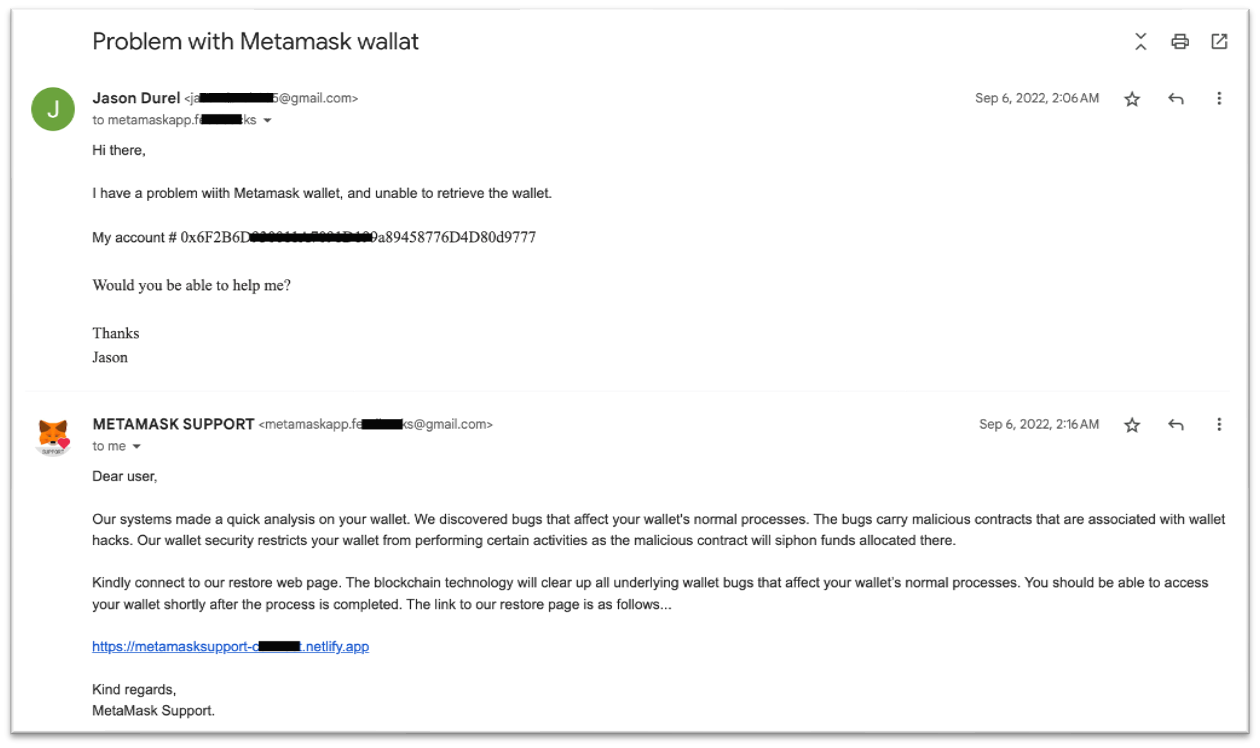}\hfill
\caption{ A sample email conversation with a scammer. In this example, we send an email asking wallet recovery support and the scammer shares a phishing link which leads to surrendering the private key phrases of the wallet.}
\label{fig:meta_mask_scammer_email}
\end{figure}

Based on the distribution of wallets in~\autoref{table:wallet_target_breakdown}, we created emails that were tailored to target the scammer corresponding to the wallet. Our emails were composed of auto-generated texts to match the wallet type. For instance, if the scammer's email alias contained the keyword "MetaMask" (metamask****@email.com), our email template generator crafted a unique email, requesting support for the ``MetaMask'' wallet. In case the scammer's email address did not mention a specific wallet type, we randomly selected a wallet from~\autoref{table:wallet_target_breakdown} and sent a generic support email. In~\autoref{fig:meta_mask_scammer_email}, we provide a sample email interaction during the system incubation period. Note that in response to our email, the scammer asked us to submit the private key to a phishing website.

\BfPara{Setup} To automate our email correspondence with scammers, we used the SendGrid API~\cite{sendgridapi}. We created an automated workflow whereby if a scammer sent an email address to our honey profiles, the system generated a support email for that email address. We created three Gmail accounts to perform the email correspondence.

\BfPara{Results} Our system generated emails to each of the 375 email addresses collected by honey profiles. We found 37 emails that encountered a delivery issue, indicating that either the email addresses were not correct or Gmail had blocked them due to abuse history. Upon closer inspection, we found that 19 of those email addresses were linked to MetaMask. Other blocked email addresses were linked to TrustWallet, Coinbase, or generic ``dev'' support. For the 338 emails that were delivered, we received a response from 74 scammers. Interestingly, we found two broad categories of scammers that responded to our emails. Scammers in the first category (27) requested private key submissions through Google Docs or URLs. Scammers in the second category requested a service fee to be paid through PayPal (22 responses) or a cryptocurrency address (25 responses). In Sec.~\tsref{sec:private_key}, we will provide more details about our follow-up interactions with scammers of both categories.

\subsection{Interactions via Instagram} 
In addition to the automated analysis, we also conducted manual interactions with scammers. We decided that manual analysis will provide a confirmation for automated analysis output. Additionally, if there were any key differences between automated and manual analysis, we could closely inspect those differences and tailor our automated approach accordingly. For manual analysis, we engaged with scammers on Instagram by collecting the Instagram handles shared by them (see~\autoref{table:scammer_scam_media_comparison_table} for details). We used the direct messaging feature for our experiment.

\BfPara{Setup} We set up four Instagram accounts for this experiment. To prevent interactions with regular Instagram users, we did not upload display pictures on our accounts. Once the accounts were set up, we conversed with scammers via manual interaction. An example script is shown in~\autoref{fig:insta_message}.

\begin{figure}[t]
\begin{theo}[Pythagoras' theorem]{thm:pythagoras}
\scriptsize
 Hi [Twitter Handle] - Thank you for reaching me on Twitter posts. I am contacting you back after seeing your comments on my tweet posts. I am in need of wallet support. I have a problem accessing my Metamask wallet address 0x41 .... aa56. I am not sure what is the problem. I need this help asap, please! Thanks in advance.
\end{theo}
\caption{Sample Instagram message.}
\label{fig:insta_message}
\end{figure}

\begin{figure}[t]
\centering
\includegraphics[width=.25\textwidth]{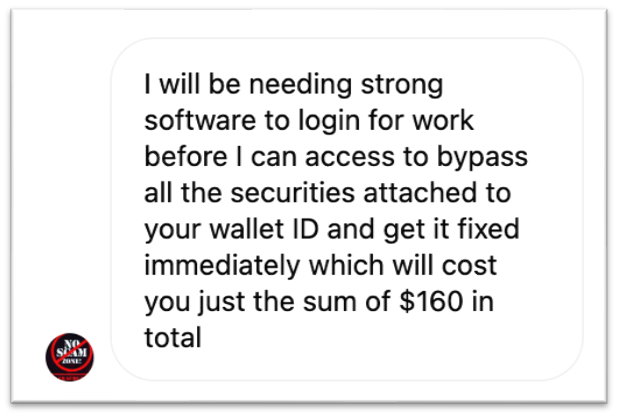}\hfill
\includegraphics[width=.218\textwidth]{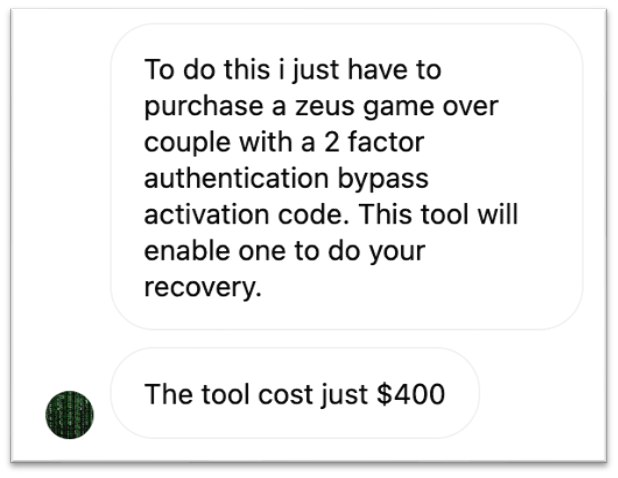}\hfill
\caption{A sample conversation we had with a scammer account on Instagram. We requested the details about the issue and the scammer responded that they required sophisticated software to diagnose the wallet issues  }
\label{fig:scammer_pay_out_examples}
\end{figure}

\BfPara{Results} We sent messages to 454 scammer accounts and received a response from 383 accounts. Through those responses, scammers requested a service fee for wallet recovery. A sample response is shown in~\autoref{fig:scammer_pay_out_examples}. We observed scammers' preferred payment modes included PayPal, cryptocurrency addresses, and gift cards. The service price ranged from \$150 to \$2,550, with a median price value of \$725. Overall, we collected 78 PayPal accounts and 89 cryptocurrency addresses. 15 accounts requested payments through \emph{CardDelivery} \cite{carddelivery} and \emph{Amazon} \cite{amazongiftcard}.

\BfPara{Key Takeaways} From our automated and manual conversations with scammers, we made the following key observations: We found no distinct difference between manual and automated interactions, highlighting that the automated approach was as effective as the manual approach and can therefore be generalized. Moreover, scammers can be categorized into two broad types based on their scam methodology. The first category of scammers requested private key submissions. We suspected that after receiving the private key, scammers would create the corresponding public key and transfer funds to their wallet addresses.
The second category of scammers requested a service fee to help with wallet recovery by citing bogus reasons. Their preferred payment methods included PayPal, cryptocurrency  addresses, and gift cards.

\section{Technical Scams: Following the Money Trail}
\label{sec:private_key}

To briefly summarize our previous analysis: We set up \htw to trap scammers and analyze their Twitter footprint. We observed that scammers quickly pivot from Twitter to other communication channels where they either request key phrase submissions or service fee deposits into their accounts. At this point, it is logical to assume that (1) if victims deposited their recovery key phrase, scammers would withdraw funds from the corresponding wallet, and (2) if victims deposited a service fee, scammers would take it and provide no further assistance. As a next step in our analysis, we decided to empirically validate those assumptions by conducting two experiments. First, we decided to create honey cryptocurrency addresses and submit their private keys to scammers. Second, we decided to share scam account details with PayPal, requesting their feedback on those accounts. In this way, if scammers used our private keys to steal money or if PayPal provided fraud confirmation, our assumptions about scam account behaviors could be validated. In the following, we provide details about our experiments. 

\subsection{Analyzing Key Phrase Theft}
\label{sec:private_key_theft}

To analyze the key phrase theft, we created 100 Ethereum \emph{honey wallet addresses} and transferred \$1.26 to each address through our MetaMask non-custodial wallet. We then released those private keys on all communication channels requested by scammers. In~\autoref{table:scammer_released_private_key_phrases}, we provide details about the key phrases released and stolen by scammers for each communication channel. 

Out of 100 released key phrases, we found that scammers stole 35 of them and moved funds from them. From those 35 wallets, funds were transferred to 26 distinct wallet addresses. The asymmetry between the number of stolen keys and the number of recipient wallet addresses indicates collusion among scammers. It is possible that scammers had created multiple identities across different communication channels. As a result, we observed a gap between the number of stolen keys and the number of recipient wallet addresses. We also received messages from 9 scammers, requesting to increase the deposit amount in our wallet addresses. The requested amount ranged from \$500 to \$3,500. There are multiple inferences that can be made regarding their requests. It is possible that they did not consider an amount of \$1.26 to be worth the stealing effort. It is also likely that scammers believed that we had completely fallen for the scam, and by requesting a higher amount, they could extract more money from us. 

\begin{table}[t]
\centering
\setlength\tabcolsep{10pt} 
\renewcommand{\arraystretch}{1.}
\scriptsize
\begin{tabular}{lcc}
\toprule
\bf{Media} & \bf{Key Phrases Sent} & \bf{Key Phrases Stolen}  \\
\midrule
Email & 25 & 8\\
Forms & 30 & 17\\
Instagram & 20 & 3\\
Telegram & 5 & 2\\
URLs & 20 & 5\\
\midrule
All & 100 & 35\\
\bottomrule
\end{tabular}
\caption{Private key phrases of unique Ethereum wallets that were released on different channels and stolen by scammers}
\label{table:scammer_released_private_key_phrases}
\end{table}

\subsection{Tracking Scam Wallets}
\label{sec:tracking_cryptocurrency_wallets}
As mentioned in Sec.~\ref{sec:email_intreactions}, we found two groups of scammers in our email interactions. The first group requested private key submissions, while the other group requested payouts to their cryptocurrency addresses or PayPal accounts. In this section, we present our findings regarding those wallet addresses and PayPal accounts. 

\subsubsection{Cryptocurrency Addresses}
We collected 136 wallet addresses shared by scammers over the communication channels. Among them, we found 72 Bitcoin and 64 Ethereum addresses. We used Bitcoin and Ethereum blockchains to monitor the transfer activity of those addresses. Below, we provide insights about each cryptocurrency.

\begin{figure}[t]
\centering
\includegraphics[width=.45\textwidth]{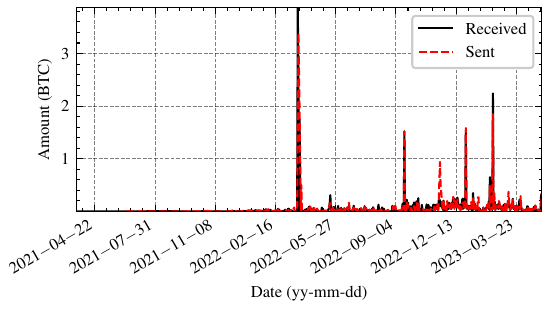}\hfill
\caption{ Amount of Bitcoins sent or received over time by addresses controlled by scammers }
\label{fig:over_all_btc}
\end{figure}

\BfPara{Bitcoin} Our Bitcoin analysis revealed that among the total of 72 addresses, 49 addresses sent or received transactions. Those 49 addresses received 3,234 transactions with a total amount of 38.40 BTC (corresponding to about \$1.117.000 at the time of writing). Those addresses also sent 2,426 transactions with a total amount of 37.74 BTC (about \$1.098.000). At the time of writing this paper, 16 addresses had a non-zero balance, and the total amount across those addresses was 0.659 BTC. In~\autoref{fig:over_all_btc}, we illustrate the transfer activity of Bitcoin addresses between 2021 and 2023. We found that incoming transactions were quickly sent to other addresses. We also found two scam addresses controlled by the same scammer that sent and received Bitcoin transactions. Using the Bitcoin blockchain, we applied the co-spent clustering heuristic and found that one address was in a large cluster of thousands of addresses, while the other address was in a small cluster of one address. We suspect that the larger cluster could be a Bitcoin exchange where a scammer hosted their wallet. 

\begin{figure}[t]
\centering
\includegraphics[width=.45\textwidth]{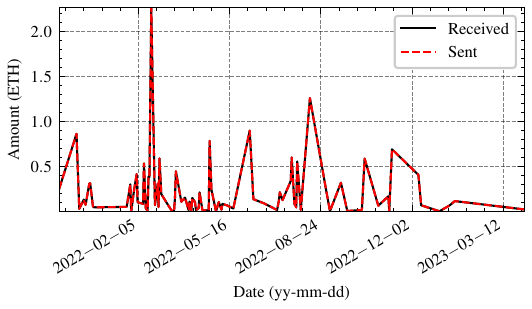}\hfill
\caption{ Amount of Ethereum sent or received over time by addresses controlled by scammers. }
\label{fig:over_all_eth}
\end{figure}

\BfPara{Ethereum} Our Ethereum analysis revealed that among the total of 64 addresses shared by scammers, 5 addresses received 80 transactions from 80 unique sender addresses. The total transferred amount was 18.9 ETH (corresponding to about \$36.000 at the time of writing) between 01-27-2022 and 05-12-2022. Moreover, among those five addresses, three addresses sent three transactions with a total value of 18.8 ETH. At the time of writing this paper, all five addresses had a positive balance. The combined value in all five addresses was 0.147 ETH. In~\autoref{fig:over_all_eth}, we show the transfer activity of Ethereum addresses, and we again observe that received payments are quickly sent to other addresses.

From the blockchain analysis, we conclude that scammers have successfully trapped victims on social media, indicated by the transfer activity on their wallet addresses. Although the transferred amount indicates that cryptocurrency-based technical support scams are prevalent, we also observe that not all scam attempts are successful. Among the total addresses collected (172), only 54 addresses (31\%) received payments. Moreover, scammers with Bitcoin addresses tend to have a better success rate than scammers with Ethereum addresses. 
\subsubsection{Analyzing PayPal Accounts}

To confirm our findings about potential scammers, we shared our data with PayPal and requested feedback. The data we shared included 101 email addresses of scammers that requested payments through PayPal. Since PayPal has recently started supporting cryptocurrency transfers, we also shared 110 wallet addresses with them to check if any PayPal user transferred cryptocurrency to those addresses. For our dataset, we requested the following insights from PayPal: (1) the number of email addresses that registered a PayPal account, (2) the geographical distribution of those accounts, (3) cryptocurrency transfers to the wallet addresses owned by scammers, and (4) confirmation of suspicion based on internal signatures collected by PayPal including affinities among scammers. PayPal agreed to provide aggregated information on scammer accounts which, is presented below.

For the email addresses we shared, PayPal confirmed fraudulent activities linked to those addresses. 97\% of the accounts were already detected and restricted by PayPal due to fraudulent activities. An additional 294 accounts were found that were linked to scammers and were restricted for involvement in similar fraudulent activities. In terms of geographical locations, among the total number of restricted accounts, 66\% registered from China, 22\% from Kenya, 5\% from the USA, and the remaining 7\% from other locations.

For the cryptocurrency wallet addresses we shared, PayPal confirmed seven transfers to three wallet addresses. Those transfers were made by four customers, indicating that four PayPal customers fell victim to technical support scams and transferred money to the wallet addresses provided by the scammers. PayPal shared that the scammer accounts we provided also transferred cryptocurrency to three unattributed wallet addresses. 

From the insights shared by PayPal, we made the following key inferences. (1) Real-world users have fallen victim to these scams. (2) Scammers use multiple modes of payment for their scams, including fiat and cryptocurrency transfers. (3) Scammers create custodial and non-custodial wallets and move cryptocurrencies between them to layer funds. (4) The scale of abuse is larger than what we have captured in our study as indicated by PayPal's detection of 294 additional accounts linked to 101 email addresses we shared with them. (5) PayPal largely detects and takes action on these scammers, thereby confirming the initial premise of our work and its merits. In summary, our tracking of cryptocurrency accounts and the feedback received from PayPal confirm our assumptions that scammers eventually steal key phrases or receive service fees without offering any technical assistance. 

\section{Scam Detection Efficacy: A Social Media Perspective}
\label{sec:social_media_robustness_and_robustness}
Our analysis so far confirms that technical support scam is a real-world problem and it is being monitored by prominent payment service providers. Therefore, it is pertinent to examine if social media platforms are also monitoring this scam and blocking the scam accounts. In this section, we conduct experiments to analyze the robustness of social media platforms in combating this scam. Based on our findings, we propose recommendations for social media platforms to improve their scam detection methodologies.

\BfPara{Experiment Setup} In order to test the scam detection efficacy of social media platforms, we collected the messages sent by scammers across all communication channels and analyzed if their messages were blocked by the hosting platform. For email addresses, we tracked email delivery failures using the SendGrid API~\cite{sendgridapi}. For Google Forms, we used the Selenium library~\cite{seleniumpython} by rendering the page in Chrome browser and checking if the page is unavailable. For Instagram, we used beautifulsoup4~\cite{beautifulsoup4} to check if the scam account was still active on the platform. For Telegram, we used the official Telegram API~\cite{telegrambotapi}, and for WhatsApp, we used the{\em Pywhatkit} library~\cite{whatsappapi} and monitored if our message to the scammer was successfully delivered. 

\BfPara{Results} In~\autoref{table:scammer_scam_media_blocked}, we report our results from the experiment. Of 6,167 total media contents linked to the scammer profiles, only  2,092 (33.93\%) were blocked by the respective platforms. We found that Twitter had the most robust detection methodology, followed by Instagram and Google forms. Surprisingly, Telegram did not act on \emph{any} scam-related content. In the following, we take a closer look at the performance of key communication channels. 

\begin{table}[t]
\centering
\setlength\tabcolsep{10pt} 
\renewcommand{\arraystretch}{1.1}
\scriptsize
\begin{tabular}{lr}
\toprule
\bf{Scam Media} & \bf{Blocked \%} \\
\midrule
Email & 8.80\\
Forms & 35.28\\
Instagram & 57.55\\
Telegram & 0.00\\
Twitter & 70.20\\
WhatsApp & 31.77\\
\midrule
All & 33.93\\
\bottomrule
\end{tabular}
\caption{Efficacy of social media platforms in blocking scam-related contents.}
\label{table:scammer_scam_media_blocked}
\end{table}

\begin{figure}[t]
\centering
            \includegraphics[width=.45\textwidth]{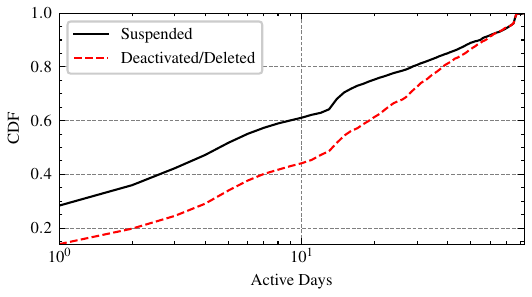}     
            \caption{Twitter scam account active days before becoming inactive - In this graph, we display the number of days Twitter accounts are alive before either being blocked by Twitter (suspended)  or the user chooses themselves to be offline via deactivation or account deletion} 
         \label{fig:suspended_deactivated_account}
\end{figure}

\subsection{Effectiveness of Twitter}
\label{sec:twitter_effectiveness}
As shown in~\autoref{table:scammer_scam_media_blocked}, Twitter outperformed other social media platforms in detecting scam-related content. In~\autoref{fig:suspended_deactivated_account}, we plot the scam accounts based on their status of being suspended or not found. We obtain that data by querying the user details Twitter API. Of the 9,149 scamming accounts that interacted with our \emph{HoneyTweet} module, Twitter suspended 70.20\%. 27.66\% of the accounts were found deleted or deactivated. Although these numbers seem somewhat reassuring, it should be noted that suspended and deactivated graph lines in~\autoref{fig:suspended_deactivated_account} converged at 90 days. In other words, these accounts were present on the platform for 90 days, which is a sufficient window to scam various customers. Note that our analysis duration was close to three months, during which we came across over 9K scammers. Therefore, we advocate for faster detection and suspension of scam accounts on Twitter.

\BfPara{Recommendation} Acknowledging the fact that Twitter's scam detection accuracy is better than of other platforms, we recommend that Twitter {i) add an interstitial page warning about the prevalence of cryptocurrency scams to the user whenever the user clicks on an external link contained in cryptocurrency-related tweets and ii) collaborate with other social media platforms to counter this type of scam. We acknowledge that there are barriers to collaboration among competing companies, however, our results show that technical support scams are emerging to be an ecosystem problem. Although Twitter takes reasonable measures to detect scam profiles, we argue that platform pivoting by scammers reduces the effectiveness of existing countermeasures. Therefore, it is pertinent that social media platforms collaborate on detecting such scams.

\subsection{Effectiveness of Gmail} Our dataset of scam email addresses contained 712 Gmail addresses. We found that those email addresses targeted three wallet types, namely MetaMask (366), TrustWallet (40), and Coinbase (11). Among those email addresses, we found that Gmail had blocked only 80 addresses. We found that 46 out of 80 blocked emails used the keyword \emph{MetaMask}, which accounted for 57.50\% of the total blocked email addresses. We also found 3 email addresses from \emph{Trust Wallet} and 1 email address from \emph{Coinbase} that were blocked by Gmail. The remaining 38.75\% (31/80) blocked email addresses belong to the category of individual names (7/31) and support groups (24/31).

\textbf{Recommendation} We recommend that Google (Gmail) monitor email aliases that masquerade various cryptocurrency wallet names and logos (Gmail profile pictures). We also recommend cross-platform sharing of suspicious cryptocurrency wallet addresses or profiles that request key phrase submission in the email. This can be trivially achieved by posting warning messages in emails that mention secret key phrase disclosure~\cite{googleabuse}. 

\subsection{Effectiveness of Submission Forms} Our dataset showed that scammers used 1,995 forms to request private key submissions as part of the fake wallet support. We observed two distinct types of forms used by fraudsters. The overall blocking rate of both form types was $\approx$35.28\% as shown in~\autoref{table:scammer_scam_media_blocked}. 

\BfPara{JotForm} Out of 159 JotForm in our dataset, 49 forms targeted Trust Wallet and 13 targeted MetaMask. All others remained cryptocurrency-agnostic and provided generic recovery support. Overall, the effectiveness blocking these scamming forms was found to be 15/159 (9.43\%). 
    
\BfPara{Google Forms} Out of 1,836 Google forms in our dataset, we found 474 forms targeting Trust Wallet, 279 forms targeting MetaMask, and 228 forms targeting Coinbase. We also found 133 forms targeting other wallets of honey tweet posts such as Binance, Badger, Exodus, and Ledger. The remaining 722 forms were found to target wallet users that were not part of the 10 wallets shown in~\autoref{table:wallet_target_breakdown}. On the blocking side, we observed that forms targeting Coinbase were predominantly blocked. Among 228 total forms, 186 (81.57\%) were blocked. The blocking rates of MetaMask and Trust Wallet were found to be 187/279 (67.025\%) and 289/424 (68.16\%), respectively. Unfortunately, the category of remaining wallets from our research including Binance, Badger, Exodus, and Ledger had a low blocking rate of 37/133 (27.81\%), which is lower than the blocking rate for three MetaMask, Coinbase, and Trust Wallet. Overall, the effectiveness of Google Forms blocking was $\approx$38.07\%. 

\textbf{Recommendation} Although Google Forms' blocking rate was higher than JotForm (9.43\%), we still consider $\approx$38.07\% block rating as inadequate. Unfortunately, based on the daily monitoring of blocked forms, the median active days since the first interaction with \htw was 16 days. Taking these limitations into account, we recommend Google Forms and JotForm to adopt some proactive scam mitigation techniques as discussed below. 

Form providers can take some unilateral actions to curb the scams instead of waiting for co-operation from Twitter as discussed in Section~\ref{sec:twitter_effectiveness}. First, \htw could serve as a tool to collect a list of abusive forms targeting continuously. If necessary, human analysts can be hired to investigate such forms similar to how they already inspect candidate phishing domains~\cite{acharyaAPE}. This is a defensive measure that email providers such as Gmail can also deploy. 

Alternatively, as a cheaper option, the form providers can unilaterally show interstitial warnings to all users who are seen to be coming from \texttt{twitter.com}. This can be done based on the \texttt{Referer} header associated with the \texttt{HTTP} request made to their servers. As of the time of writing this paper, \texttt{twitter.com} has set its \texttt{Referrer-Policy} header to the common default value of \texttt{strict-origin-when-cross-origin} which allows the form providers to be able to obtain this origin information on most browsers.
\section{Discussion}
\label{sec:discussion}
 
\BfPara{Responsible Disclosure} We have disclosed our findings to social media platforms including Twitter, Google, Instagram, Telegram, and WhatsApp. Our report plan included scam accounts, their characterization, and our detection techniques. We believe that our disclosure encourages collaboration among these social media platforms to mitigate cryptocurrency-based technical support scams that are shaping up to be a cross-platform ecosystem problem. We received a positive review and acknowledgment from Google. Furthermore, we shared our findings with PayPal such that the payment provider can investigate these scams. 

\BfPara{Ethical Considerations} We upheld ethical standards while conducting our experiments. For instance, after setting up \htw, we abstained from manual interactions during data collection to avoid any bias. Moreover, our experimental interactions only were carried to accounts that engaged with our honey tweets. Additionally, as outlined in Sec.~\tsref{ssec:analysis_module}, we took conservative measures to remove potential false positives from our dataset, thereby ensuring fine-grained detection of scam accounts. Additionally, we ensured that our tweet frequency also remained within the ethical realms. Our system tweeted four tweets per hour from four different Twitter handles. In total, our systems generated 16 tweets per hour, which is significantly below the maximum allowed threshold set by Twitter. 

Since this is a deception-based study with possible retribution attacks for the authors, we omitted debriefing and after-study consent procedures as safety measures. The IRBs for all institutions involved in this work agreed with this methodology. 

\BfPara{Limitations} In the experiment involving \emph{honey wallet addresses}, it is possible that scammers auctioned our key phrases to other attackers in the wild. It is also possible that due to a low balance, scammers transferred cryptocurrency from our addresses to other benign wallets. Although we acknowledge these scenarios as our potential limitations, they still largely support our observations regarding scam monetization and private key theft enabled by technical support scammers. 
\section{Related work}
\label{sec:related_work}

In this paper, we take a closer look at scammers that perform technical support scams targeting the users of social media. 
To the best of our knowledge, we are the first to employ automated baiting techniques to detect cryptocurrency scammers on Twitter. The work that is most related to ours is in the form of scam-baiting via automated emails~\cite{cambiaso2023scamming, chen2022active} and not social media. Cambiaso et. al~\cite{cambiaso2023scamming} used Chat-GPT for engaging scammers in automatized and pointless communications with the goal to waste scammers' time and resources. Chen et. al~\cite{chen2022active} also performed scam-baiting by playing the roles of victims via email responses. 

Though understanding Twitter's effectiveness in blocking abusive accounts has been a widely studied topic~\cite{suspendedThoms11, grier2010spam,thomas2013trafficking}, none of the prior work focused on baiting cryptocurrency scammers. Profiling scammers provides insights into techniques and tricks performed by scammers. The work that was most related to profiling visitors is carried by authors of~\cite{phishprint,zhang2021crawlphish,oest2019phishfarm} but rather focused on profiling and evading techniques in client-side phishing in the web domain URLs. 

With regards to previous work in cryptocurrency scams, the work from Li et al.~\cite{xigao2023doublenothing} studied cryptocurrency giveaways via Certificate Transparency logs and tracked stolen funds using public blockchain logs. Xia et al.~\cite{9493255} clustered the cryptocurrency-related phishing and scam web content to identify a typology of advance fee and phishing. Phillips and Wilder~\cite{ phillips2020tracing} created a taxonomy of cryptocurrency scams. In the area of mobile applications, the work from Hong et al.~\cite{hong2021mobilescams} performed a study of scam distribution on mobile gambling and revealed permissions misuse from phone apps leading to a share of scam messages and hosted channels performing a cryptocurrency fraud transfer. The research from Tu et al.~\cite{tu2016sok} provided a study of the telephone spam ecosystem via automated spamming phone calls.

In the last few years, technical support scams have inspired researchers to dive deep and understand the ecosystem that scammers perform in the wild. Gupta et al.~\cite{perisci2018} studied the technical support scams that targeted Twitter users by scraping posts and analyzing the phone numbers associated with them. Miramirkhani et al.~\cite{miramirkhani2017NDSS} also studied phone-based technical support scams by scraping scam websites and manually baiting the scammers. The work from Srinivasan et al.~\cite{srinivasa2018WWW} provides a study of these scams distributed via parked domains and abused sponsored advertisements.

All of the prior work relied on setting up a crawler for empirically finding the relevant scams in the wild. In contrast to the previous work, scammers would perform a wild search to find and interact with our system voluntarily. Our honey tweets are tailored to bait scammers. Thus, our system \htw is designed as a \emph{honeypot} for baiting fake technical support scammers on social media.  
\section{Conclusion and Future Work}
\label{sec:conclusion}

In this paper, we present the first systematic analysis of crypto\-cur\-ren\-cy-based technical support scams. We develop the \htw framework to lure over 9K scammers on Twitter. We track those scammers as they pivot from Twitter to other platforms, and we deploy interactive frameworks to bait them into revealing their payment information. We categorize scammers based on their payment preferences and uncover techniques used by scammers to extort victims. To confirm our findings about scammers' extortion techniques, we deploy {\em honey wallet addresses} and partner with PayPal to analyze the money trail. Our analysis on both fronts produces consistent results and exposes the end-to-end lifecycle of such scams. Consolidating our analysis across various vantage points, we propose recommendations to mitigate cryptocurrency-based technical support scams. 

Our measurements and analysis open several directions to foster future work in this domain. Our work provides a foundation for baiting crypto\-cur\-ren\-cy-based technical support scammers on social media. Acknowledging the fact that such scams are evolving fast as criminals adopt new techniques, our analysis represents a subset of technical support scam attacks. Leveraging the techniques highlighted in this work, our framework can be extended to analyze other emerging fraud types in the ecosystem.

\textbf{Acknowledgements}
We would like to thank Efrén López and Xinyi Xu for exploring other forms of attacks. This work was funded by the German Federal Ministry of Education and Research (grant 16KIS1900 ``UbiTrans''). This material is also based in part upon work supported by the National Science Foundation (NSF) under grant No. CNS-2126655. Finally, the project has also been partially financed by the European Union—NextGenerationEU (National Sustainable Mobility Center CN00000023, Italian Ministry of University and Research Decree n. 1033—17/06/2022, Spoke 10).
\bibliographystyle{ieeetr}
\bibliography{strings,bib}

\appendix
\section{Appendix}\label{sec:set-diff-dodis}

\subsection{HoneyTweet Design Rationale}
\label{sec:honeytweet_design_rationale}
This system was inspired by coincidental scam attempts experienced by the authors on Twitter. As a small separate motivating experiment, the authors created a new Twitter account (with 0 followers) and posted a two-sentence Tweet with random words that included the word “Metamask” and “help”. Note however that the tweet was not seeking any help regarding Metamask. Regardless, the tweet attracted 23 scam replies in about 4 minutes. Removing the word “help” however did not attract as much attention from the scammers. On the other hand, changing the one two-sentence Tweet to a single sentence still solicited scam replies.

The above mini-experiment demonstrates that while our Tweet structure is not sensitive to the number of sentences and semantics, it likely benefits from the use of some words that reflect the need for user support. We considered these observations when designing our \textsc{HoneyTweet} structure.

To choose crypto-currency wallets, we selected the top 10 popular wallets based on an online source. Note that we first performed this in 2022 but unfortunately, lost access to this original link. However, a similar web page that lists 8 of the 10 wallets we considered is linked here~\cite{top_wallets}.

\subsection{Inclusion/Exclusion Criteria For Tweet Filtering}
\label{sec:tweet_filtering}
For exclusion, we applied various automated strategies to ensure that our dataset does not contain benign users and posts. We provide additional detail in the below bulleted points.
\begin{itemize}
    \item \textbf{Exclude Twitter Handle of Official Wallet, Exchanges and Coins} As stated in Sec.~\tsref{ssec:analysis_module}, we automated a script that collected 56 exchanges and 8,538 cryptocurrency coins from CoinMarketCap~\cite{coinmarketcapall}. We compiled 100 popular cryptocurrency wallets based on CoinCapMarket~\cite{coinmarketcapall} and online searches~\cite{top_wallets}. For each of the cryptocurrency wallets, exchanges, and coins, we further collected their Twitter handle, domain, and email addresses. Any interacting account's Twitter handle that matches to official wallet, exchanges, and coins Twitter handle, was excluded from our dataset.
    \item \textbf{Account Features} Using Twitter's user detail API ~\cite{twitteruserdetailapi}, we excluded any Twitter handle that contained user details with the account verified status. We acknowledge that this is rather a conservative filtering, and might have potentially excluded scammers.
    \item \textbf{Legitimate Email and Links} Our automated strategy removed any account that displays links to any legitimate and well-known wallet service. For this, we simply ensured the 2LD domain name presence in the email and URLs. For example for Trust Wallet, we check emails matching \emph{**@trustwallet.com} and URLs matching \emph{**.trustwallet.com/*}.
    \item \textbf{Platform Pivoting} In our posts, we observed scammers often tried to pivot their interaction to external platforms such as Instagram, Telegram, and WhatsApp. However, this was not the case for benign user and official wallet interaction. Official wallets usually ask the user to contact them via their official wallet email or domain. 
    \item \textbf{Payment Requests and Private Key Phrases} One of the key factors that differentiate scammers vs non-scammers is that scammers ask me) some form of payment to be made as part of fake technical support or ii) ask private key phrases from the users to be released via email or direct message. Any posts or communication that shows some form of payment requests or private key phrases were marked as potential scammers. The filtering process yielded “potential scam” accounts that were forwarded to PayPal and the feedback we received confirmed that our methodology was consistently capturing scam accounts.
    \item \textbf{Additional Filtering} We ensured any posts from the user that do not contain email, links, pivoting mechanisms, payment requests, and private key phrases were marked as benign. This filtering may have potentially excluded scammers.
\end{itemize}

\subsection{Evaluation of HoneyTweet Data}
\label{sec:evaluation}
In Sec.~\tsref{ssec:analysis_module}, we described a filtering process to weed out Tweets that might potentially be false positives. We acknowledge that the approach may lead to concerns regarding process completeness and false positives after applying message filtering. To address those concerns, we reiterate that one of the important filtration criteria is to only focus on messages that steer the victim away from public Twitter messages to other communication channels as discussed in the paper. We also note that these messages (see \autoref{fig:scam_media}) betray the pattern of ``honeyness'' to any benign human who stumbles upon our accounts as the same 4 Twitter accounts that we periodically generated through our honey profiles. make. Both of these approaches significantly reduce the likelihood that any legitimate well-meaning human would manually respond to our Honey Tweets (while simultaneously directing us to an alternative communication channel such as email/third-party forms). 

At the same time, we acknowledge that there is a possibility that there might exist some legitimate third-party service that offers cryptocurrency wallet support services, although we are not aware of such legitimate services. Note that we already filtered out any messages from official wallet Twitter accounts (such as MetaMask), which sporadically post automated replies to our HoneyTweets warning us to not fall for these scams, as shown in Figure~\ref{fig:fp_screen_shot}. Therefore, we attempted to systematically verify the ground truth of our post-filtration Tweet messages. Admittedly, verifying the ground truth of Tweet messages in isolation is a challenging process as merely ~\emph{the offer of help via a secondary communication channel} is not completely indicative of an active scam. Thus, we needed more context that mandated active interaction with the potential scammer via the advertised communication channel. This process is difficult in many cases as these communication channels have a limited lifetime or the interactions involve a significant manual effort. For example, given that more than 97\% of the Twitter accounts that interacted with us ended up becoming inactive (Table~\ref{table:over_all_tweet_interact}), it is nearly impossible to conduct a post-mortem ground truth analysis for these accounts. 

Given these challenges, we pursued a best-effort approach for Tweet data evaluation based on two popular communication channels: forms and emails. For this task, we considered our collection of 1076 form URLs and 375 email addresses, each of which can be directly associated with 58\% and 53\% of all Twitter accounts that interacted textually (via replies/quoted tweets) with our HoneyTweets. We randomly sampled 100 forms and 100 email addresses from our dataset and conducted a manual investigation of the collected data.

\subsubsection{Forms} We visited all 100 forms and took screenshots using a Python-based crawler implementing the Selenium framework. Manual analysis showed that none of the 100 forms are benign: 48 were blocked due to ToS violations; 2 were deleted; 49 forms were active and attempting to steal Wallet key phrases. Interestingly, there was 1 tech-support form that did not ask for a key phrase but had a field for the email address. Our email communication eventually resulted in an attempted key phrase request, thereby revealing a new triple-platform segmented attack example ($\text{Twitter} \rightarrow \text{Forms} \rightarrow \text{Email}$).

\subsubsection{Emails} For the 100 email addresses, we observed that 83 of them were impersonating official cryptocurrency services such as MetaMask/TrustWallet (Example: \texttt{metamask***.us@gmail.com}), and 1 was an e-mail address that asked for a key phrase in earlier experiments. No definitive ground truth can be inferred about the remaining 16 addresses but their naming pattern of generic ``crypto-recovery'' services was seen in prior malicious cases (see Sec.~\ref{sec:email_intreactions}).

\subsection{Clustering of Scam Channels}
\label{sec:clustering_scammer_channel}

In this section, we further dive deep into scammer's interacted \emph{replies} and \emph{quoted} tweets with our \htw system. For each of the scam communication channels, we perform a pattern match (Email, Form, Instagram, Telegram, WhatsApp, and Twitter DM) of the contact method that scammers ask potential victims to contact. We then create a cluster based on the scam contact method. If a given contact method from scam channels has at least two scam account interactions, we label the given contact method as a cluster of a given scam channel. For example, if two or more scammers use the same email address \emph{metamask**1234@gmail.com} as a contact method in any of the tweet posts asking potential victims to contact them back via email, we count the email address as one cluster. This definition applies to other channels such as Form, Instagram, Telegram, Twitter DM, and WhatsApp. In Table~\ref{table:scammer_channel_clustering}, we provide an overview of each cluster. The first column consists of all six scam communication channels and, as the last row, their sum as \emph{Distinct (All)}. The second column \emph{Total Clusters} shows the total number of clusters found in each category of scam channels. We provide distinct tweet interactions by \emph{Replies} and \emph{Quoted} tweets in columns three and four, whereas the fifth and sixth columns provide the distinct interacted tweet posts and all posted text respectively. The scam accounts that belong to the category of the interacted cluster are shown in the seventh column. Columns 8, 9, 10, and 11 provide the cluster size evaluation in terms of minimum, median, 90th percentile, and max size. The final column, \emph{Median Seen Diff} is a median time difference (days) value between the scammer's first and last interaction with our \htw system. We highlight some of the text-based clustering data insights below. 

\textbf{Cluster Constituents.} Out of 3098 communication channels found in Honey Profiles, 2319 (74.85\%) communication channels belong to the cluster. More than 90\% of contact methods from Telegram (245/259) and Twitter DM (662/731) belong to the scam cluster whereas the contact methods from Email (324/375) and WhatsApp (90/106) were over 80\% belonging to scam cluster. The remaining two communication channels belonging to the clusters from Instagram and Form contribute to 75.68\% (417/551) and 53.99\% (581/1076) respectively.
 
\textbf{Cluster Life Span.} Our results show that the Email cluster median life span seen in Tweet interaction was the highest (8 days). The rest of the channels, except for Form (4 days), show a single-day campaign. This signifies that scammers are more likely to create newer handles to avoid being blocked by Twitter. Our clustering life span data is further validated by the account suspension and deletion from fig.~\ref{fig:suspended_deactivated_account} which shows Twitter's effectiveness in blocking 60\% of scammer's handles within 10 days. In Fig.~\ref{fig:clusters}, we further provide three graphical analyses. The first graph (a) provides cumulative clusters first seen over the experimental time. In the second graph (b) we provide clusters actively interacting on each day with the \htw system, and the third graph (c) provides the cumulative interaction difference between scam accounts from cluster and non-cluster (singleton) datasets.
 
\textbf{Shared Scam Accounts.} The intersection of scamming accounts among multiple clusters showed that scammers use a combination of one or more types of scam channels while interacting with potential victims. We provide a detailed overview of scamming accounts found shared among multiple clusters of different scam channels in Table~\ref{table:heatmap_shared_campaign_scammers}. Among the distinct 7870 scammers found in all six communication channels of clusters, more than 50\% of the scammers perform one or more types of campaign scams in Form (4642/7870), Twitter DM (4154/7870) and Instagram (4031/7870). The text highlighted in grey color shows the highest preference of scammers in performing a second scam channel distribution while interacting with our \htw profiles. For example, more than 50\% scammers from the Email and Form clusters also performed Twitter DM-based scam campaigns. Similarly, scammers from Instagram, Telegram, and Twitter DM are more likely to perform Form-based scam campaigns as the second preference in luring potential victims to contact them back. In summary, this analysis provides a preference-based behavioral understanding of scammers across multiple scam communication channels. We suspect scammers involved in multiple scam channel usages allow potential victims to have more options in contacting them back, which is likely an advantage to scammers in yielding a better conversion rate as part of scam monetization.

\textbf{Prolific Scam Accounts.} We conducted a deeper analysis of all the 2319 clusters of Twitter accounts as described in Table~\ref{table:scammer_channel_clustering} to find evidence for any prolific groups. For this, we grouped the 2319 clusters in an agglomerative fashion if they ever happen to post messages containing the same social platform identifiers (such as email address, Instagram handle, or form URL). Interestingly, the 2319 clusters agglomerated into 43 groups. Of these groups, there exists one prolific group that accounted for as many as 7751 out of all the 7870 scam Twitter accounts depicted in Table~\ref{table:scammer_channel_clustering}. Furthermore, our analysis showed that this group alone accounted for 33782 out of all 47368 (71.31\%) interactions that were made with our honey tweets.

\begin{figure*}
    \centering
    \subfigure[]{\includegraphics[width=0.30\textwidth]{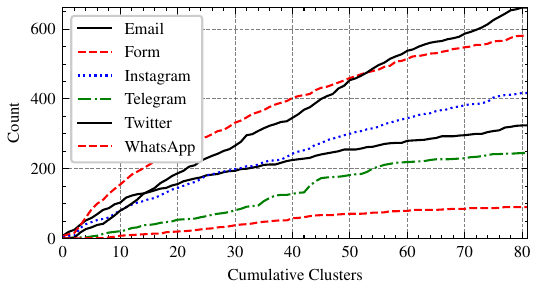}} 
    \subfigure[]{\includegraphics[width=0.30\textwidth]{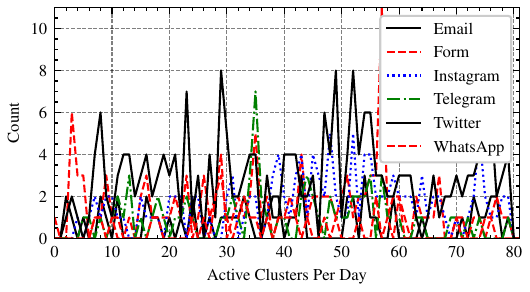}}
    \subfigure[]{\includegraphics[width=0.30\textwidth]{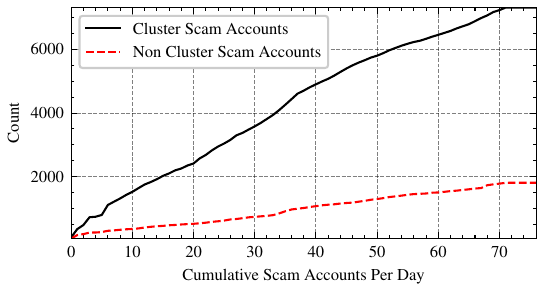}}
    \caption{The left graph (a) shows the cumulative clusters seen during our experiment time from all six communication channels. The middle graph (b) displays the active clusters on a given day throughout the experiment time. The right graph (c) shows the cumulative scam accounts that belong to the category of clusters and non-clusters. }
    \label{fig:clusters}
\end{figure*}

\begin{table}[t]
\scalebox{0.75}{
\setlength\tabcolsep{2.5pt} 
\renewcommand{\arraystretch}{1.2}
\begin{tabular}{lcccccc}
\toprule
\bf{Channels} & \bf{Email} & \bf{Form} & \bf{Instagram} & \bf{Telegram} & \bf{Twitter DM} & \bf{WhatsApp} \\ 
\midrule
Email & 3507 & 2381 & 1816 & 1044 &{\colorbox{lightgray}{1758}} & 500 \\
\hline
Form & 2381 & 4642 & 2419 & 1413 & {\colorbox{lightgray}{2875}} & 639 \\
\hline
Instagram & 1816 & {\colorbox{lightgray}{2419}} & 4031 & 1291 & 2391 & 755 \\
\hline
Telegram & 1044 & {\colorbox{lightgray}{1413}} & 1291 & 2218 & 1306 & 432\\
\hline
Twitter DM & 1758 & {\colorbox{lightgray}{2875}} & 2391 & 1306 & 4154 & 694 \\
\hline
WhatsApp & 500 & 639 & {\colorbox{lightgray}{755}} & 432 & 694 & 1019 \\
\bottomrule
\end{tabular}
}
\caption{Distribution of scammers performing shared campaigns among multiple communication channels as part of pivoting victims. The gray highlight shows each of the channel's preferred second-highest communication channels on each row.}
\label{table:heatmap_shared_campaign_scammers}
\end{table}

\begin{table*}[t]
\setlength\tabcolsep{3pt} 
\renewcommand{\arraystretch}{1.1}
\scriptsize
\begin{tabular}{lccccccccccc} 
\toprule
\midrule
\bf{Scam}  & \bf{Total} & \bf{Distinct} & \bf{Distinct} & \bf{Distinct} & \bf{All} & \bf{Total} & \bf{Min} & \bf{Median} & \bf{90Pct} & \bf{Max} & \bf{Median} \\
\bf{Channels} & \bf{Clusters} & \bf{Replies TwtID} & \bf{Quoted TwtID} & \bf{TweetID} & \bf{Text} & \bf{Scammers}& \bf{Clust. Size} & \bf{Clust. Size} & \bf{Clust. Size} & \bf{Clust. Size} & \bf{Seen Diff}\\
\midrule
Email & 324 & 1733 & 1945 & 2532 & 11524 & 3507 & 2 & 21 & 118 & 649 & 8 \\
\hline
Form & 581 & 3482 & 1577 & 3591 & 13729 & 4642 & 2 & 16 & 78 & 532 & 4 \\
\hline
Instagram & 417 & 1572 & 632 & 2006 & 9215 & 4031 & 2 & 10 & 51 & 317 & 1 \\
\hline
Telegram & 245 & 737 & 96 & 821 & 3906 & 2218 & 2 & 8 & 31 & 200 & 1\\
\hline
Twitter DM & 662 & 2608 & 304 & 2693 & 9545 & 4154 & 2 & 9 & 32 & 283 & 1\\
\hline
WhatsApp & 90 & 180 & 125 & 269 & 1693 & 1019 & 2 & 8 & 30 & 163 & 1\\
\hline
Distinct (All) & 2319 & 7326 & 3883 & 8077 & 24838 & 7870 & 2 & 11 & 98 & 649 & 1 \\
\hline
\end{tabular}
\caption{Clustering of scammers by channels found based on interaction with \emph{Replies} and \emph{Quoted} tweets}
\label{table:scammer_channel_clustering}
\end{table*}

\subsection{Profile Image Clustering Hyperparameters and Visualizations}
\label{sec:profile_clustering}
In the following, we provide details on the hyperparameters and visualization clusters that we use in~\ref{sec:scammer_profile_pictures}

\textbf{Clustering Hyperparameters}\label{sec:hyperparameters}
For both UMAP and DBSCAN, we systematically evaluated the clustering performance using a combination of hyperparameters from the specified ranges. We employed a common evaluation metric, i.e., silhouette score~\cite{Shahapure2020ClusterQA}, and visual inspection of resulting clusters to assess the quality and validity of the clustering outcomes. This hyperparameter tuning aims to optimize the performance of the clustering pipeline and achieve meaningful and reliable results. For this purpose, we considered a wide range of hyperparameter configurations. Specifically, for UMAP, we let the two most influential hyperparameters, i.e.,  \texttt{n\_neighbors} and the \texttt{n\_components} vary in the intervals $[15, 105]$ and $[2, 128]$ respectively. Regarding DBSCAN, we let the \texttt{min\_cluster\_size} and \texttt{min\_dist} vary in the intervals $[10, 50]$  and $[1e-02, 1]$ respectively. The resulting investigation involved $2,500$ configurations of these hyperparameters, identifying the configuration \texttt{n\_neighbors}=85, \texttt{n\_components}=2, \texttt{min\_dist}=1, and \texttt{min\_cluster\_size}=20 as the most reliable for our clustering pipeline.

\textbf{Visualizing clusters of scammers}
{In Fig.~\ref{fig:scammers_cluster_1} and Fig.~\ref{fig:scammers_cluster_2} we show, for each cluster we identified in Sec.~\ref{sec:scammer_profile_pictures}, a subset of $50$ scammer profile pictures. Complementary, in Fig.~\ref{fig:miscellaneous}, we illustrate samples coming from the Miscellaneous cluster. Notably, the content inside the NFTs, Male, Female, Wallet, Tech Support, and Default Twitter Profile clusters we identified is cohesive and coherent with our assigned label. On the other hand, the Miscellaneous cluster (only ~1\% of the data) contains a mixture of profile pictures that have been considered anomalous by our clustering algorithms. }

\begin{figure*}[!htbp]
\centering
\includegraphics[clip, trim=2.5cm 1.5cm 2.5cm 1.1cm,width=0.3\textwidth]{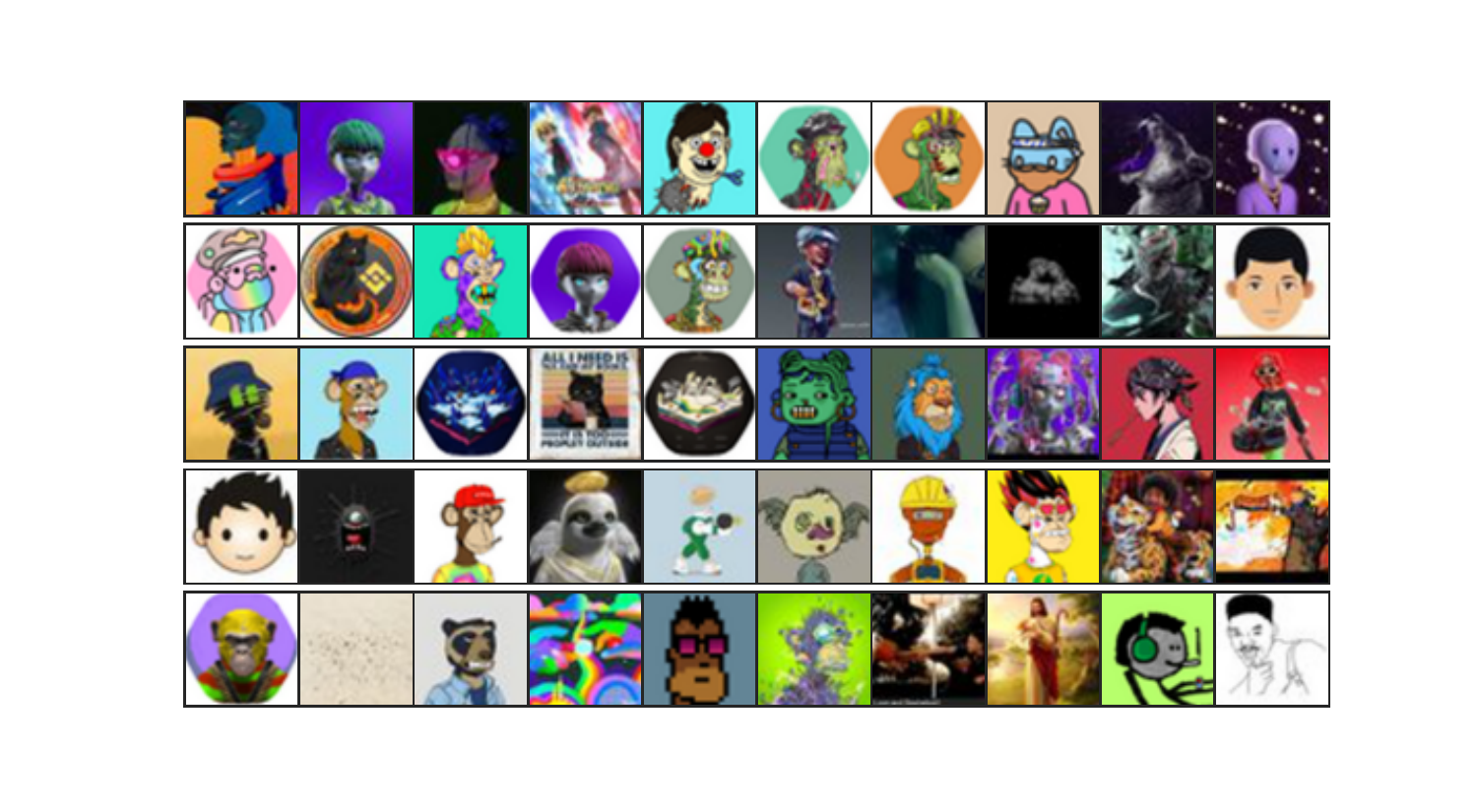}\hfill
\includegraphics[clip, trim=2.5cm 1.5cm 2.5cm 1.1cm,width=0.3\textwidth]{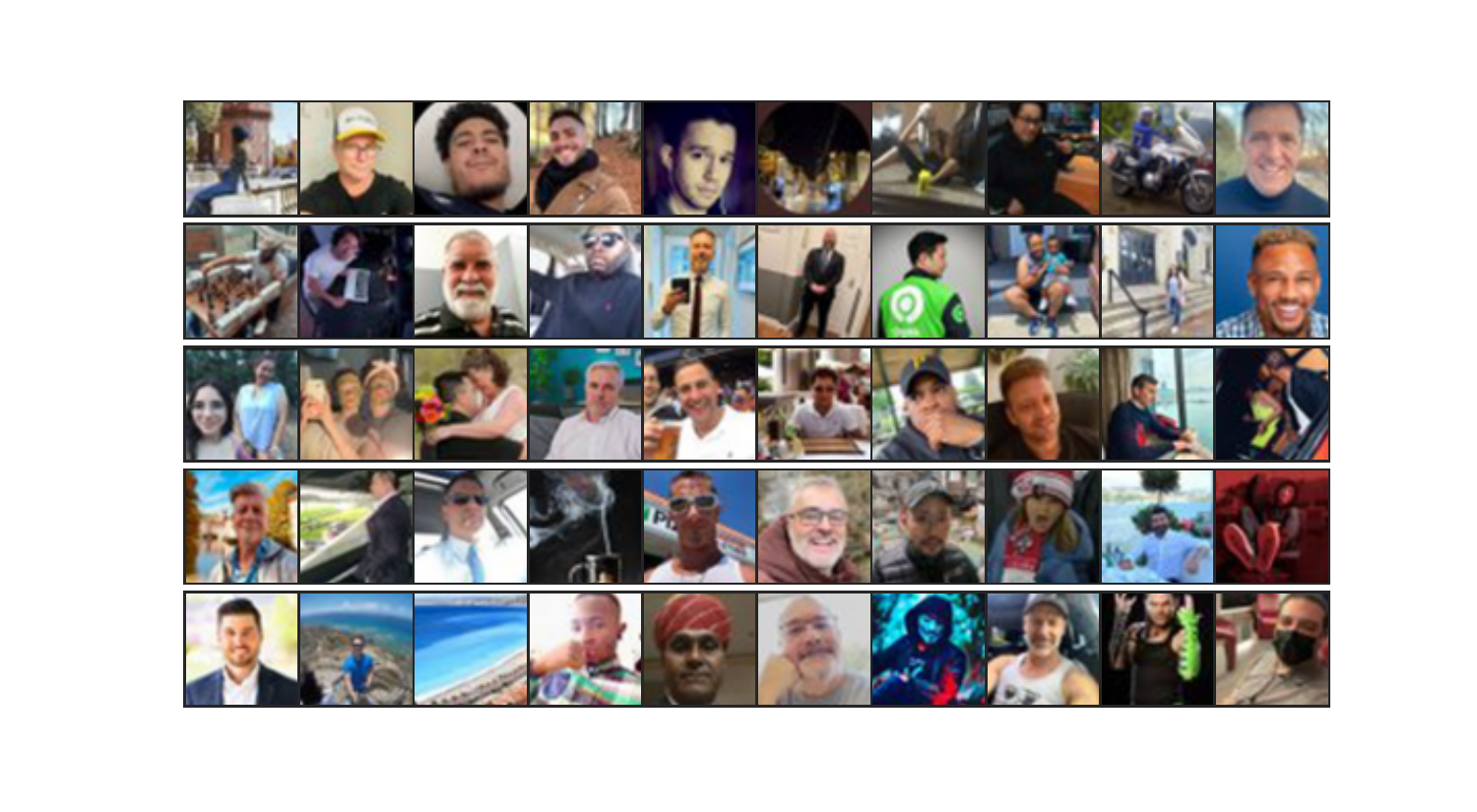}\hfill
\includegraphics[clip, trim=2.5cm 1.5cm 2.5cm 1.1cm,width=0.3\textwidth]{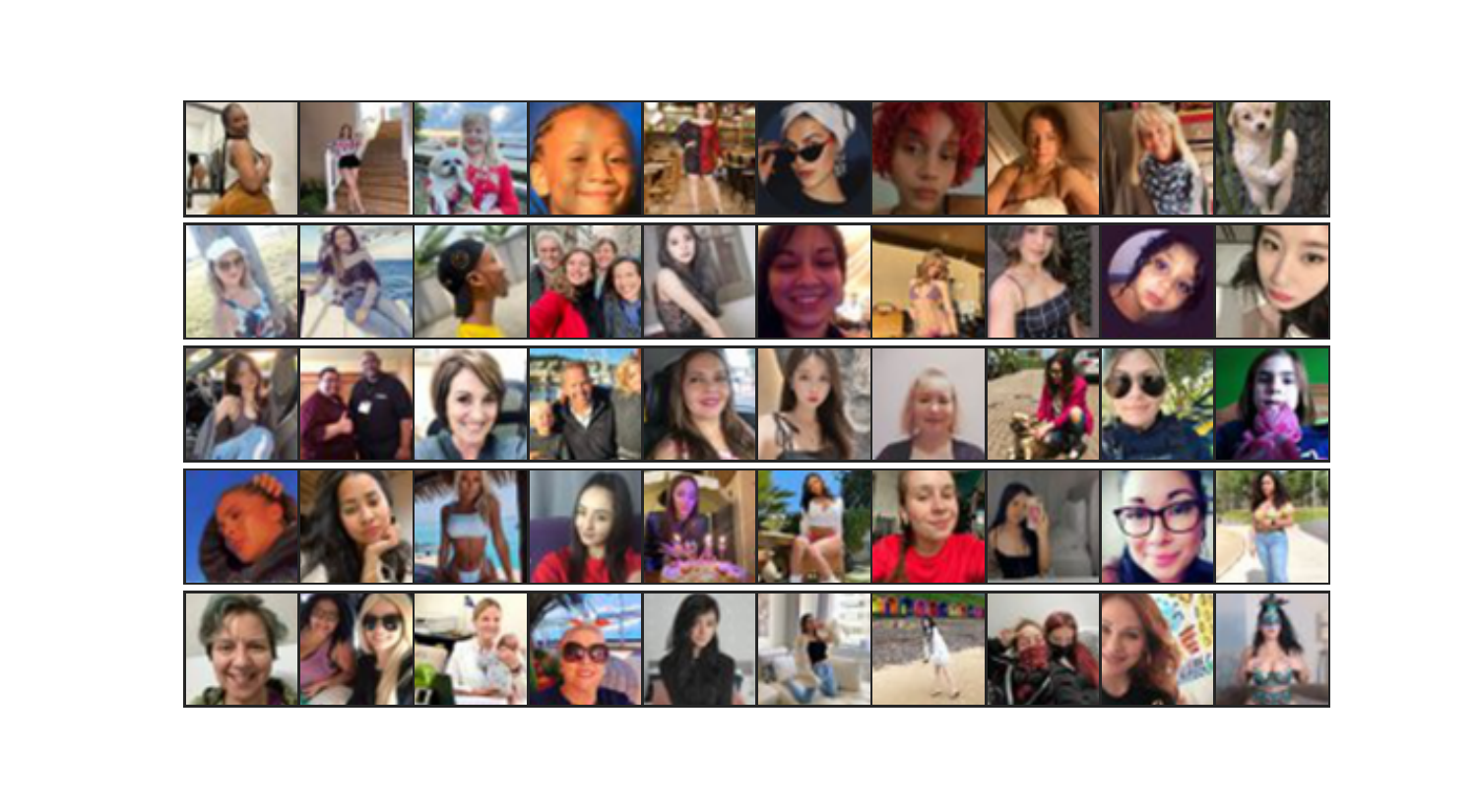}
\caption{Visualization of each of 50 samples from NFTs (left), Male (middle), and Female (right) clusters of scammers.}
\label{fig:scammers_cluster_1}
\end{figure*}
\begin{figure*}[!htbp]
\centering
\includegraphics[clip, trim=2.5cm 1.5cm 2.5cm 1.1cm,width=0.33\textwidth]{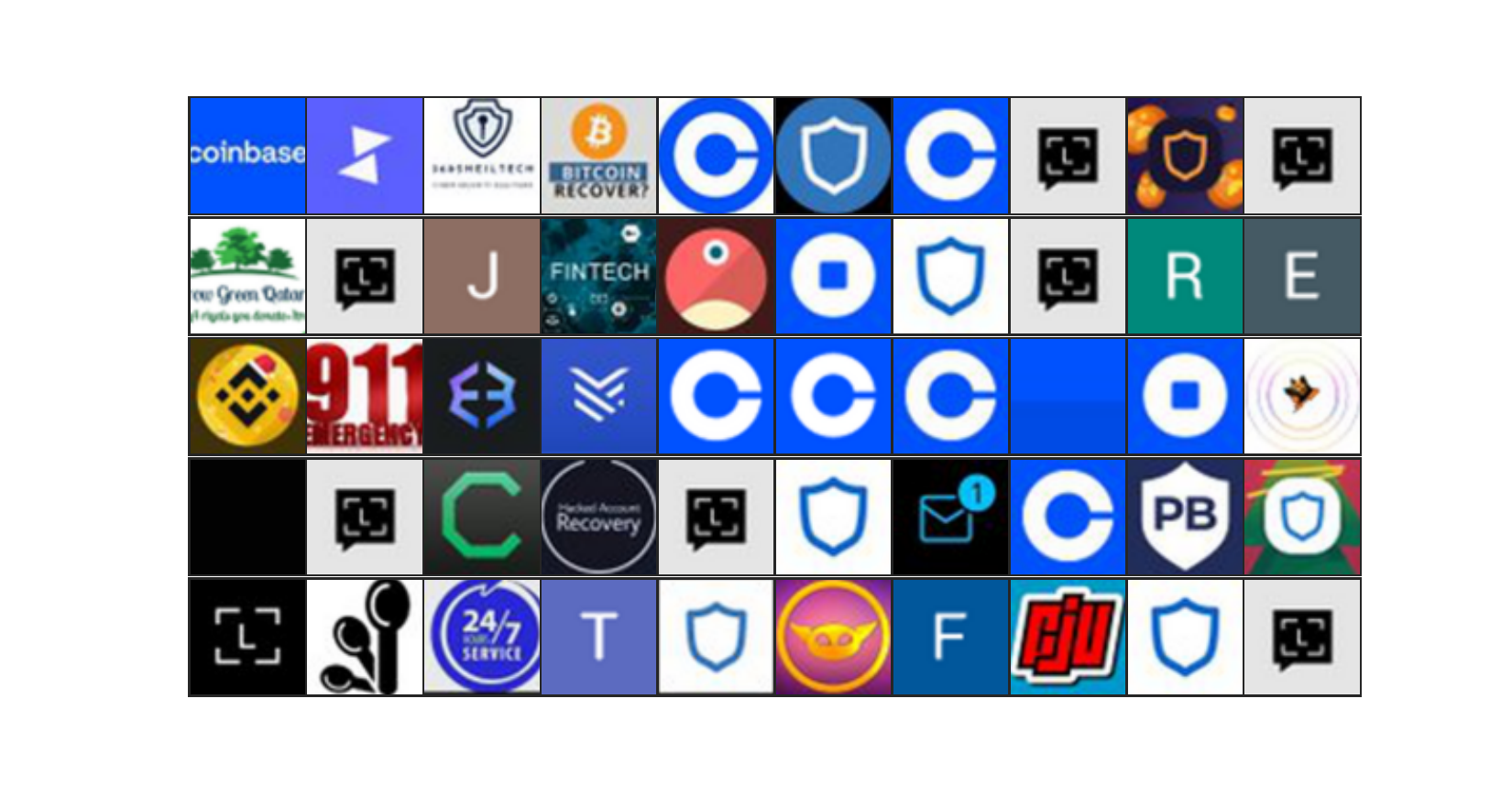}\hfill
\includegraphics[clip, trim=2.5cm 1.5cm 2.5cm 1.1cm,width=0.33\textwidth]{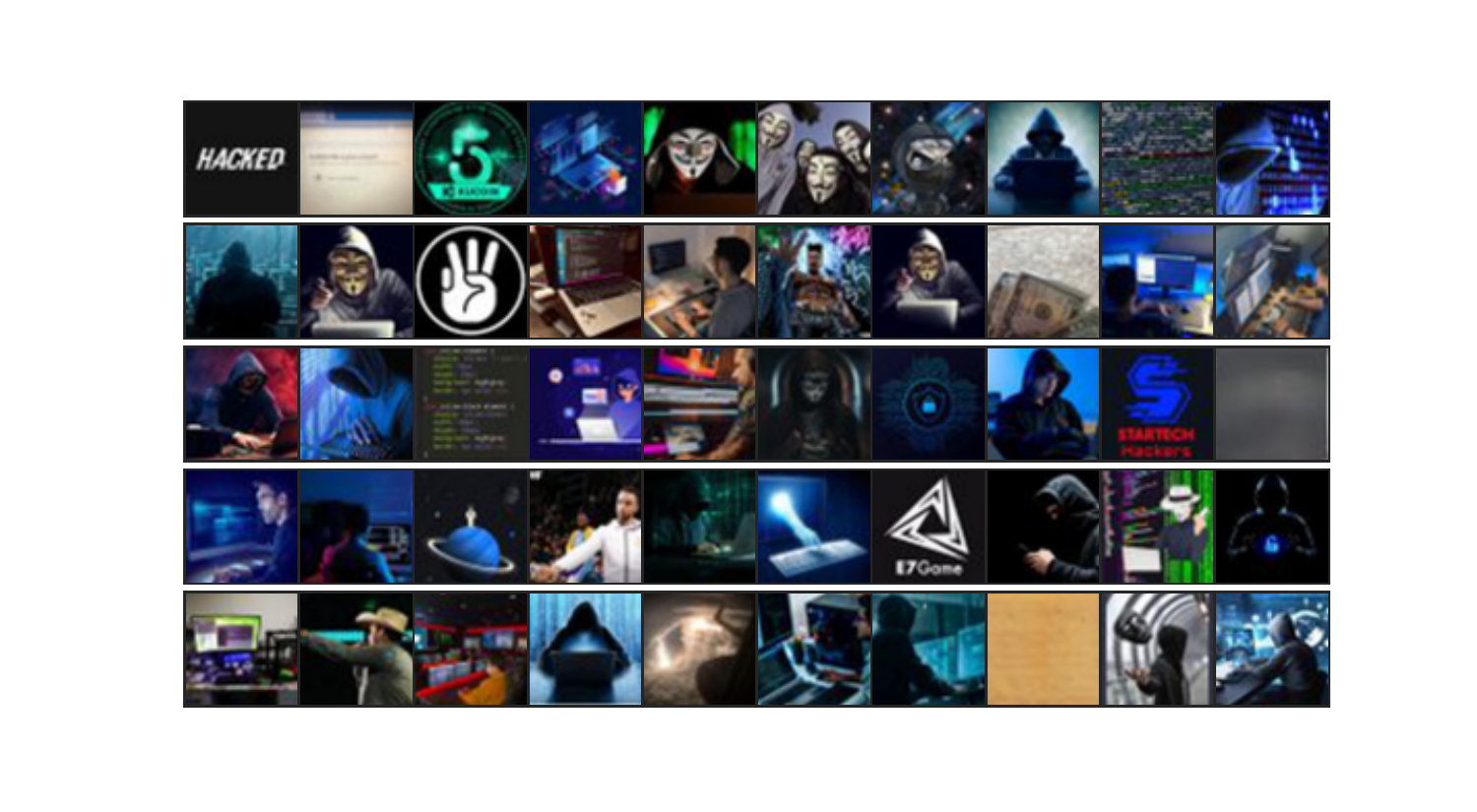}\hfill
\includegraphics[clip, trim=2.5cm 1.5cm 2.5cm 1.1cm,width=0.33\textwidth]{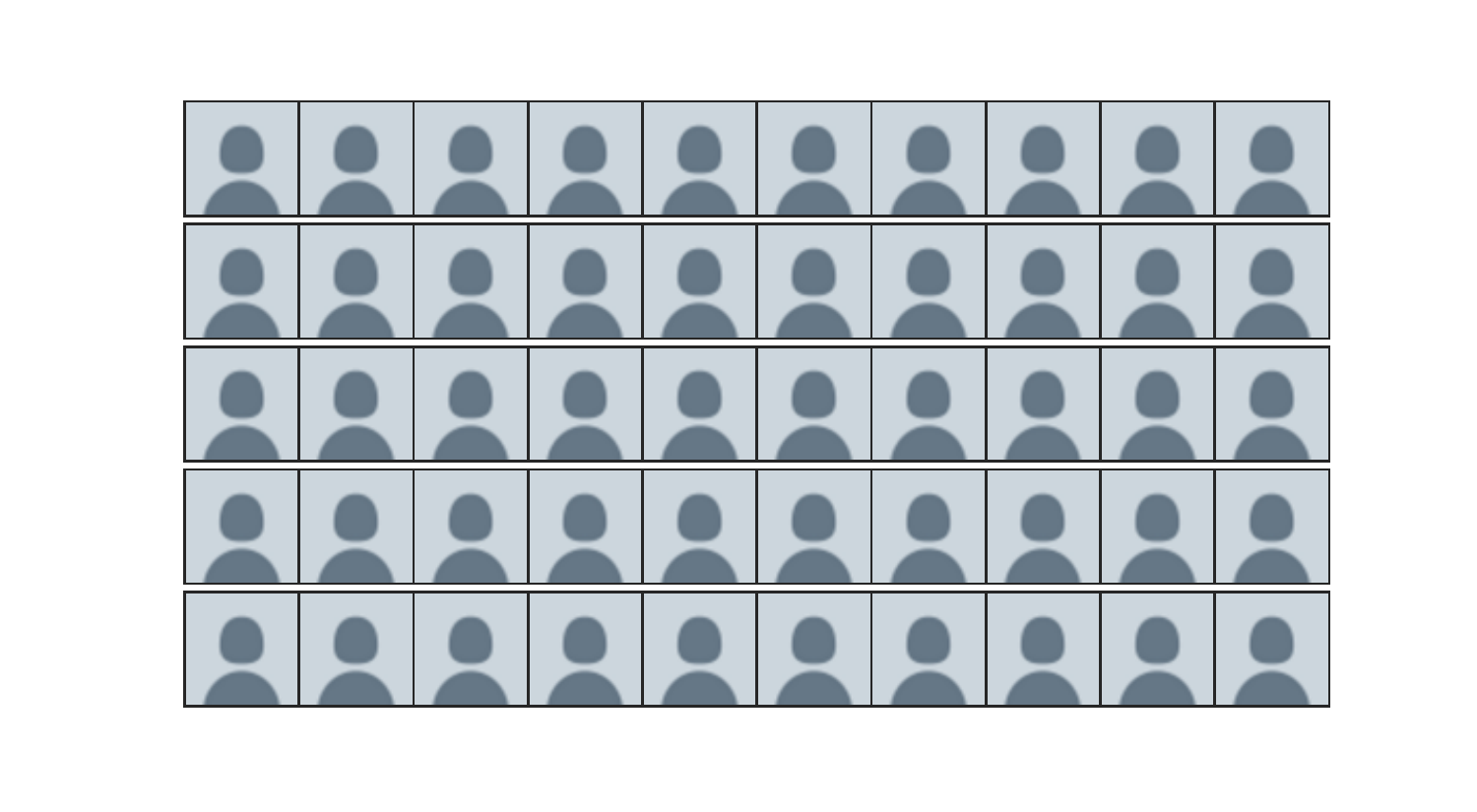}
\caption{Visualization of each of 50 samples from Wallet (left), Tech Support (middle) and Default Twitter Profile (right) clusters of scammers.}
\label{fig:scammers_cluster_2}
\end{figure*}
\begin{figure*}[!htbp]
\centering
\includegraphics[clip, trim=2.5cm 1.5cm 2.5cm 1.1cm,width=0.33\textwidth]{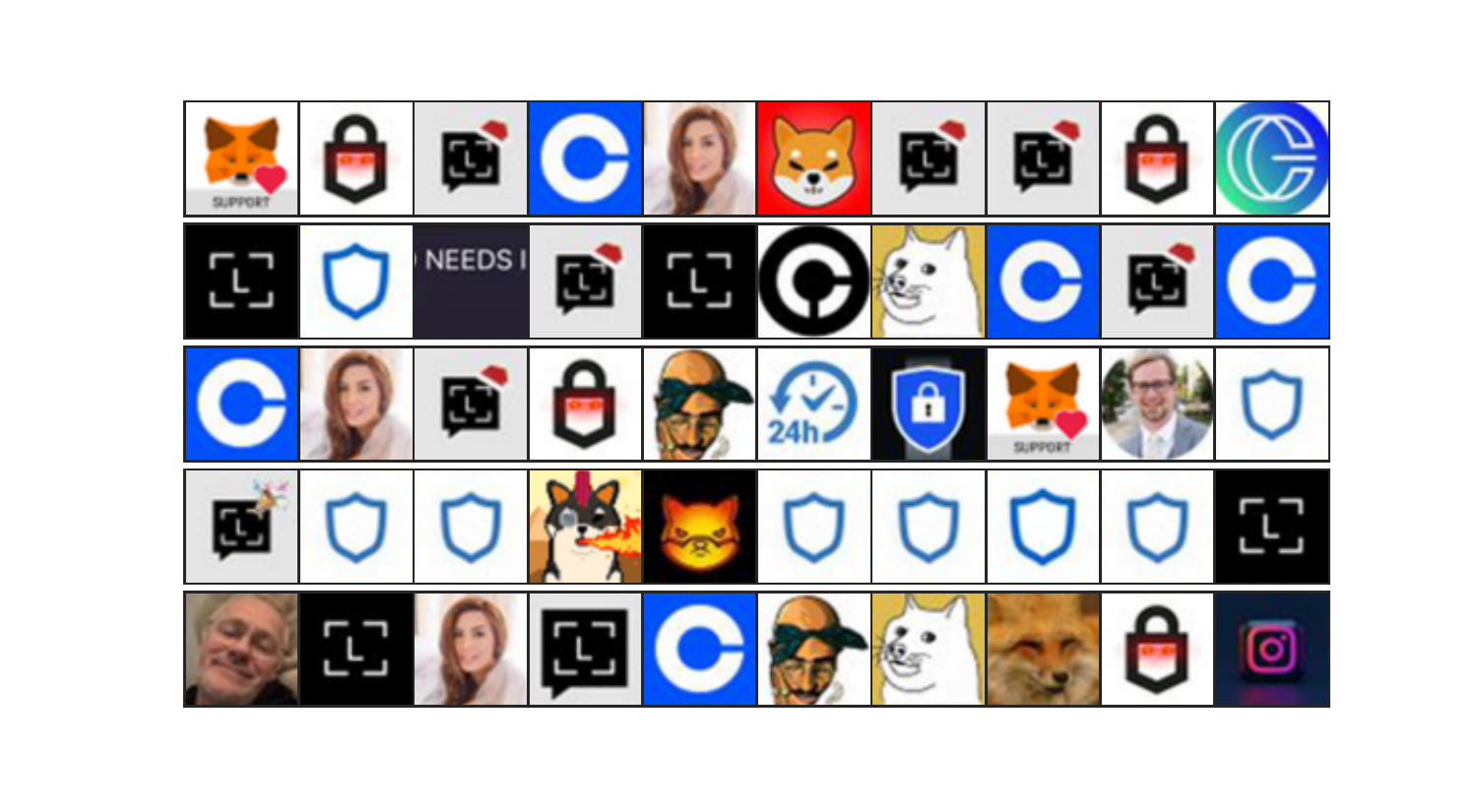}
\caption{Visualization of each of 50 samples from Miscellaneous clusters of scammers.}
\label{fig:miscellaneous}
\end{figure*}

\subsection{Phishing/Other Attacks Differentiation}
In this work, we focus on scams in which the attackers attempt to gain unauthorized access to decentralized cryptocurrency wallets (e.g., MetaMask, TrustWallet), which are typically protected by a key phrase made up of a randomly generated set of words. Note that there are two subtle but key differences between the authentication mechanisms for the wallets and credentials based (such as banking and e-commerce) that are often a target of phishing websites.
\begin{enumerate}[leftmargin=*]
    \item Decentralized cryptocurrency wallets, by design, \emph{do not have an identifying non-secret User ID} with them, unlike a phishing target website. Unfortunately, all wallet users’ authentication flow only includes the use of the secret key phrase~\cite{decentralize1, decentralize2}. From the point of view of a potential victim who is troubleshooting their access to a wallet, this could become problematic. As we will see later, this makes a user more prone to divulging their key phrases, even in cross-platform channels such as emails and web-based forms, as this is the~\emph{only} information they use to identify their wallets. On the other hand, we do not know of any study yet that documented in-the-wild successful usage of such channels for stealing website passwords.

    \item The cryptocurrency wallets we consider in this paper are cryptographically bound to the key phrases. By design, it is impossible to provide common website security mechanisms such as resetting credentials (i.e., key phrases), providing two-factor authentication ~\cite{twofactor}, or checking concurrent/recent logins in these wallets. This provides successful attackers an~\emph{immediate} and \emph{irreversible} (permanently lasting) chance to steal funds from victim accounts which is a much more lucrative proposition to attackers than regular phishing attacks.
\end{enumerate}
Although the impact of stealing key phrases is different from a credentials-based phishing attack, the attack vector itself (e.g., replies to tweets) is still an effective mechanism in finding old or emergent spam/scam ecosystems. 

\newpage 

\appendices 

\section{Meta-Review}

The following meta-review was prepared by the program committee for the 2024
IEEE Symposium on Security and Privacy (S\&P) as part of the review process as
detailed in the call for papers.

\subsection{Summary}
This paper investigates cryptocurrency-based technical support scams conducted on Twitter designed to steal cryptocurrency from accounts that post for private-key or wallet-related help on the platform. The paper conducts a thorough analysis of the lifecycle of these scams, starting from their own generated HoneyTweets (e.g., tweets designed to get engagement from scammers) through both automated and manual interactions with scammers, and ultimately all the way through the wallets and money channels that scammers ultimately request. The paper builds a system, HoneyTweet, that conducts all these steps and the authors present the results in a clear and compelling narrative detailing the scam ecosystem, ultimately also finding that such scams do appear to be effective in the wild (roughly \$1.1M USD in BTC stolen at the time of writing.) The paper concludes with some recommendations for platforms and highlights the cross-platform nature of the issue moving forward.

\subsection{Scientific Contributions}
\begin{itemize}
\item Independent Confirmation of Important Results with Limited Prior Research
\item Addresses a Long-Known Issue
\item Creates a New Tool to Enable Future Science
\item Provides a Valuable Step Forward in an Established Field
\end{itemize}

\subsection{Reasons for Acceptance}
\begin{enumerate}
\item Provides a new tool that creates “HoneyTweets” to interact with cryptocurrency scammers online
\item Timely study that focuses on an issue that continues to plague online interactions (spam / scams)
\item Overall interesting results with potential for impact
\end{enumerate}

\end{document}